\documentclass[]{article}
\pdfoutput=1
\usepackage{amsmath}
\usepackage{graphicx}
\usepackage{amssymb}

\makeatletter
\newcommand{\lyxaddress}[1]{
\par {\raggedright #1
\vspace{1.4em}
\noindent\par}
}
\usepackage{geometry}
\usepackage{cite}
\begin{document}

\title{Superlattice Patterns in the Complex Ginzburg-Landau Equation with
Multi-Resonant Forcing}

\author{Jessica M. Conway$^{1}$ and Hermann Riecke$^{1,2}$}

\maketitle

\lyxaddress{$^{1}$Engineering Sciences and Applied Mathematics, Northwestern
University, Evanston, IL 60208, USA}

\lyxaddress{$^{2}$Northwestern Institute on Complex Systems, Northwestern University,
Evanston, IL 60208, USA}

\begin{abstract}
Motivated by the rich variety of complex patterns observed on the
surface of fluid layers that are vibrated at multiple frequencies,
we investigate the effect of such resonant forcing on systems undergoing
a Hopf bifurcation to spatially homogeneous oscillations. We use an
extension of the complex Ginzburg-Landau equation that systematically
captures weak forcing functions with a spectrum consisting of frequencies
close to the 1:1-, the 1:2-, and the 1:3-resonance. By slowly modulating
the amplitude of the 1:2-forcing component  we render the bifurcation
to subharmonic patterns supercritical despite the quadratic interaction
introduced by the 1:3-forcing. Our weakly nonlinear analysis shows
that quite generally the forcing function can be tuned such that resonant
triad interactions with weakly damped modes stabilize subharmonic
patterns comprised of four or five Fourier modes, which are similar
to quasi-patterns with 4-fold and 5-fold rotational symmetry, respectively.
Using direct simulations of the extended complex Ginzburg-Landau equation
we confirm our weakly nonlinear analysis. In simulations domains of
different complex patterns compete with each other on a slow time
scale. As expected from energy arguments, with increasing strength
of the triad interaction the more complex patterns eventually win
out against the simpler patterns. We characterize these ordering dynamics
using the spectral entropy of the patterns. For system parameters
reported for experiments on the oscillatory Belousov-Zhabotinsky reaction
we explicitly show that the forcing parameters can be tuned such that
4-mode patterns are the preferred patterns.

\end{abstract}

\section{Introduction}

Complex but ordered spatio-temporal patterns, characterized by multiple
length scales, have been observed and studied in a range of systems.
In particular in the Faraday system, in which a thin layer of fluid
is vertically vibrated leading to patterns on the surface of the fluid,
various kinds of superlattice patterns and quasi-patterns have been
found experimentally depending on the frequency content of the forcing
function \cite{ChAl92,EdFa93,EdFa94,KuGo96,KuPi98,ArFi00,ArFi02,WeBi03,DiUm06}.
 Related complex patterns have also been observed in optical systems
\cite{HeWe99}, in vertically vibrated fluid convection, where they
arise from the competition of different instability mechanisms \cite{RoPe05},
and on the surface of ferrofluids driven by time-periodic magnetic
fields, where they are due to spatial period-doubling \cite{KoLe02}. 

From a purely spatial point of view the quasi-patterns as well as the
superlattice patterns observed in the above forced dissipative systems
are closely related to quasi-crystals obtained in thermodynamic equilibrium
\cite{ShBl84}. There the significance of damped, resonating modes
for the stabilization of quasi-crystals has been recognized \cite{MeTr85,NePo93}.
Recently, two-dimensional quasi-crystals have been found in soft-matter
systems \cite{ZeUn04}; their dodecagonal symmetry has been discussed
in the context of a Swift-Hohenberg-type model \cite{LiDi07} that
had previously been introduced to get insight into Faraday patterns
\cite{LiPe97}. 

The temporal aspect introduced by the forcing of non-equilibrium systems
greatly increases the richness of patterns observed. To wit, contributing
to the great variety of  complex patterns observed in the Faraday
system is the ability to control in detail the temporal wave form
of the vibration's forcing function \cite{EdFa94,KuPi98,ArFi02,SiTo00,PoTo04}.
This allows extensive tuning of the interaction between plane waves
of different orientation, which can in turn stabilize superlattice
patterns and quasi-patterns. Two main stabilization mechanisms have
been identified, both of which exploit the presence of weakly damped
resonating modes. In one case the competition between plane-wave modes
of different orientation is suppressed for a relatively narrow range
in the angle subtended by the competing modes \cite{SiTo00,PoSi03,PoTo03,PoTo04,RuSi07}.
In the other case the self-coupling of each mode is strongly enhanced,
rendering the competitive coupling between the modes effectively weak
over a considerable range in the angle \cite{ZhVi96,ChVi97,ZhVi97,ZhVi97a,ZhVi98,ChVi99,RuSi07}
and allowing patterns comprised of multiple modes to become stable
\cite{MaNe89}. The latter mechanism captures qualitatively the scenario
envisioned for the stabilization of `turbulent crystals' \cite{NePo93}.
We will focus on this second mechanism and will make use of the fact
that the temporal forcing introduces an additional symmetry that can
be exploited to select or suppress certain types of patterns. 

In this paper we investigate, motivated by the richness of patterns
observed in the Faraday system, whether such complex spatio-temporal
patterns are also accessible in systems undergoing a Hopf bifurcation
to spatially homogeneous oscillations. These systems constitute a
class that differs from those in which quasi-patterns have
been observed previously and they may offer the potential for additional
complexity through the interaction between the spontaneous oscillation
arising from the Hopf bifurcation and the external forcing. Chemical
oscillations like those observed in the Belousov-Zhabotinsky reaction
are a classic example of such a system. 

In the absence of any temporal forcing the Hopf bifurcations we have
in mind lead to spatially homogeneous oscillations or long-wave traveling
waves, which may break up to form spirals or more complex chaotic
states (e.g. \cite{ChMa96,OuFl96,BeOu97,ArKr02,MaRi06a}). With temporal
forcing, which in the case of the Belousov-Zhabotinsky reaction can
be achieved by time-dependent illumination to exploit the photosensitivity
of the reaction, the oscillations can become locked to the forcing
and - depending on the ratio between the Hopf frequency and the forcing
frequency - different types of patterns can arise \cite{CoEm92a,CoFr94,PeOu97,LiBe00,LiHa00,LiHa04}.
Thus, forcing near twice the Hopf frequency can lead to competition
between two types of domains differing in their temporal phase or
to labyrinthine patterns. For a frequency ratio of 1:3, labyrinthine
patterns, spirals, or competing domains have been observed. Ordered
 patterns with multiple length scales have been observed experimentally
in chemical systems so far only when a \emph{spatially periodic} illumination
mask was applied to initialize the pattern \cite{DoBe01,BeYa03}.
We focus on the case of \emph{spatially uniform} illumination.

The organization of the paper is as follows. In Sec.\ref{sec:CGLE}
we  extend the complex Ginzburg-Landau equation (CGLE)  to include
the terms that describe the external forcing at various frequencies.
The derivation of this equation for the Brusselator, which is a simple
model for chemical oscillations, is sketched in the Appendix. In Sec.\ref{sec:Linear-Stability-Analysis}
we present a linear stability analysis of the equation to find the
onset of standing waves that are phase-locked to the driving. In particular
we focus on the vicinity of the codimension-2 point at which the subharmonic
and the harmonic standing waves bifurcate simultaneously. To determine
the stability of the desired subharmonic patterns we derive in Sec.\ref{sec:Weakly-Nonlinear-Analysis}
the corresponding amplitude equations by performing a weakly nonlinear
analysis of the CGLE, and then use energy arguments to guide us in
terms of the relative stability of various pattern comprised of different
numbers of modes. In order to  confirm our predictions for the pattern
selection, we perform in Sec.\ref{sec:Results} numerical simulations
of the CGLE in small and large domains, and characterize the temporal
evolution of patterns by using a spectral pattern entropy. A brief
account of these results has been published previously in \cite{CoRi07}.

\section{The Complex Ginzburg-Landau Equation\label{sec:CGLE}}

We are interested in the formation of complex patterns in systems
that undergo a Hopf bifurcation at vanishing wave number and that
are forced at frequencies near integer multiples of the Hopf frequency.
Near the Hopf bifurcation and for weak forcing such  systems can be
described by a suitably extended complex Ginzburg-Landau equation
(CGLE), the form of which can be derived using symmetry arguments. 

In the absence of forcing the complex amplitude $A(t)$ satisfies
the usual CGLE, \begin{equation}
\frac{\partial A}{\partial t}=a_{2}A+a_{3}\Delta A+a_{4}A|A|^{2},\label{eq:nothing}\end{equation}
 where $a_{i}\in\mathbb{C},i=1\dots4$ and $\Delta$ is the Laplacian
in two dimensions \cite{ArKr02}. The CGLE exhibits the normal-form
symmetry $T_{\tau}:\, A\rightarrow Ae^{i\omega\tau}$ reflecting the
invariance of the original system under translations of the time $\tilde{t}$
by an arbitrary amount $\tau$, $T_{\tau}:\tilde{t}\rightarrow\tilde{t}+\tau$.
Here $\omega$ is the Hopf frequency. The slow time $t$ is given
by $t=\delta^{2}\tilde{t}$, $0<\delta\ll1$. 

The extension of the CGLE that describes strongly resonant multi-frequency
forcing of the form $F=f_{1}e^{i\omega\tilde{t}}+f_{2}e^{2i\omega\tilde{t}}+f_{3}e^{3i\omega\tilde{t}}+c.c.$
can be obtained by considering the analogous center-manifold reduction
of the extended dynamical system in which the forcing amplitudes $f_{1}$,
$f_{2}$, and $f_{3}$ are considered as dynamical variables that
vary on the slow time scale $t$. Under time translations $T_{\tau}$
they transform as $f_{1}\rightarrow f_{1}e^{i\omega\tau},\, f_{2}\rightarrow f_{2}e^{2i\omega\tau},\, f_{3}\rightarrow f_{3}e^{3i\omega\tau}$
(e.g. \cite{RiCr88}). To cubic order in $A$ the most general equation
that is equivariant under $T_{\tau}$ is then given by \begin{equation}
\frac{\partial A}{\partial t}=a_{1}+a_{2}A+a_{3}\Delta A+a_{4}A|A|^{2}+a_{5}\bar{A}+a_{6}\bar{A}^{2},\label{eq:cgle_early}\end{equation}
 where $a_{1}=b_{11}f_{1}+b_{12}\bar{f}_{2}f_{3}$, $a_{2}=b_{21}+b_{22}|f_{3}|^{2}$,
$a_{5}=b_{51}f_{2}$, $a_{6}=b_{61}f_{3}$. The $b_{ij}$ are O(1)
complex coefficients. For all terms to appear at the same order in
the final amplitude equations, we use the scaling $A(t)=O(\delta)$,
$f_{1}=O(\delta^{3})$, $f_{2}=O(\delta^{2})$, and $f_{3}=O(\delta)$.
Spatial coordinates are scaled as $(x,y)=\delta(\tilde{x},\tilde{y})$,
where $(\tilde{x},\tilde{y})$ are the original spatial coordinates
of the system.

The forcing terms $f_{j}$ satisfy decoupled evolution equations on
their own. In the simplest case this evolution expresses a detuning
$\nu_{j}$ of the forcing $f_{j}$ from the respective resonance and
the $f_{j}$ satisfy \begin{equation}
\frac{df_{j}}{dt}=i\nu_{j}f_{j},\qquad j=1\ldots3.\label{eq:detuning}\end{equation}
 In general, the detuning introduces time dependence into (\ref{eq:cgle_early}). 

In a previous study \cite{CoRi07a} we investigated the simpler case
in which the forcing parameter $f_{1}$ satisfies $b_{11}f_{1}=-b_{12}\bar{f_{2}}f_{3}$
and the detuning parameter $\nu_{3}$ satisfies $\nu_{3}=3\nu_{2}/2$,
which allowed us to make all coefficients of the extended CGLE (\ref{eq:cgle_early})
time-independent by a transformation $A\rightarrow Ae^{i\nu_{2}t/2}$.
For that case we found that the possibly stable superhexagon pattern,
a superlattice pattern, arises in a transcritical bifurcation. We
showed that the amplitudes on the upper branch of the transcritical
bifurcation are $O(1)$ and that therefore a weakly nonlinear analysis
taken to cubic order is insufficient to describe these patterns. The
bifurcation is transcritical because the quadratic term in the CGLE
introduces also a quadratic term in the amplitude equations describing
the competing Fourier modes. To eliminate the latter quadratic term
without losing the resonant triad interaction we choose in the following
the forcing near twice the Hopf frequency to be quasi-periodic, \begin{equation}
f_{2}=f_{21}+f_{22}\qquad\mbox{with}\qquad\frac{df_{21}}{dt}=i\nu_{21}f_{21},\quad\frac{df_{22}}{dt}=i\nu_{22}f_{22}.\label{eq:quasi-periodic}\end{equation}
The difference between the two detunings $\nu_{21}$ and $\nu_{22}$
introduces then a periodic time dependence in  (\ref{eq:cgle_early}).
To obtain the desired patterns, we exploit the spatiotemporal resonances
induced by the time dependence and focus on patterns that are subharmonic
in time. The quadratic interaction of two subharmonic modes induces
a harmonic mode and does not generate a quadratic term in the amplitude
equation for the subharmonic mode. Consequently, we expect a  pitch-fork
bifurcation. 

Considering the quasi-periodic forcing (\ref{eq:quasi-periodic})
we simplify (\ref{eq:cgle_early}) by absorbing the time dependence
of $f_{21}$, $e^{i\nu_{21}t}$, into $A$ through the transformation
$A\rightarrow Ae^{i\nu_{21}t/2}$. Further, we write $a_{3}=1+i\beta$
by rescaling the spatial coordinates and $a_{4}=-(1+i\alpha)$ by
rescaling $A$. We focus on the case of a supercritical Hopf bifurcation
and choose the real part of $a_{4}$ to be negative. We now introduce
restrictions on the forcing $f_{j}$ for the purpose of making the
analysis manageable at the expense of some generality. To eliminate
the homogeneous term $a_{1}$ we choose $b_{11}f_{1}=-b_{12}(\bar{f}_{21}+\bar{f}_{22})f_{3}$.
To remove the time dependence from the $\bar{A}^{2}$ term we choose
$\nu_{3}=3\nu_{21}/2$. This yields\begin{equation}
\frac{\partial A}{\partial t}=(\mu+i\sigma)A+(1+i\beta)\Delta A-(1+i\alpha)A|A|^{2}+(\gamma_{1}+\gamma_{2}e^{i(\nu_{22}-\nu_{21})t})\bar{A}+\eta\bar{A}^{2}.\label{eq:cgle_medium}\end{equation}
 Here the growth rate $\mu$ is modified by the forcing function $|f_{3}|^{2}$,
$\sigma$ is a linear function of the detuning $\nu_{21}$, and $\eta$
is a complex linear function of $|f_{3}|$. We rewrite $\eta$ in
magnitude and phase, $\eta=\rho e^{i\Phi}$, and write $\gamma_{1}=\gamma\cos(\chi)$
and $\gamma_{2}=\gamma\sin(\chi)$ with $\chi$ characterizing the
relative forcing strengths. Finally we set $\nu\equiv\nu_{22}-\nu_{21}$,
to get the version of the complex Ginzburg-Landau equation (CGLE)
that we will investigate in the following, 

\begin{equation}
\frac{\partial A}{\partial t}=(\mu+i\sigma)A+(1+i\beta)\Delta A-(1+i\alpha)A|A|^{2}+\gamma\left(\cos(\chi)+\sin(\chi)e^{i\nu t}\right)\bar{A}+\rho e^{i\Phi}\bar{A}^{2}.\label{eq:cgle_final}\end{equation}

In this paper we stay below the Hopf bifurcation and focus on the
case $\mu<0$. As in the Faraday system, in the absence of forcing,
the only solution is then the basic state $A=0$.

\section{Linear Stability Analysis}

\label{sec:Linear-Stability-Analysis}

To obtain the onset of standing waves that are phase-locked to the
driving we consider the stability of the solution $A=0$ of (\ref{eq:cgle_final})
\cite{CoFr94}. Linearizing (\ref{eq:cgle_final}) about $A=0$ and
splitting the equations into real and imaginary parts by setting $A=A_{r}+iA_{i}$,
we obtain linear partial differential equations for $A_{r}$ and $A_{i}$.
Since these equations are time-periodic with period $2\pi/\nu$ we
use Floquet theory in the same way as described in \cite{KuTu94}.
The solutions to the linear problem in Floquet form are \begin{equation}
\left(\begin{array}{c}
A_{r}\\
A_{i}\end{array}\right)=e^{(\hat{\delta}+i\zeta)\nu t}\sum_{n=-\infty}^{\infty}\left(\begin{array}{c}
X_{n}\\
Y_{n}\end{array}\right)e^{in\nu t+ikx}+c.c.\label{eq:LinAnsatz}\end{equation}
 Here $e^{(\hat{\delta}+i\zeta)t}$ is the Floquet multiplier. We
restrict $\zeta$ to two values corresponding to a harmonic response
($\zeta=0$) or to a subharmonic response ($\zeta=1/2$).
To obtain the neutral curve for the standing waves, the ansatz (\ref{eq:LinAnsatz})
is substituted into the linearized system with $\hat{\delta}=0$,
yielding the infinite-dimensional system of equations\begin{align}
\sum_{n=-\infty}^{\infty}e^{i(n+\zeta)\nu t} & \left\{ \left(\begin{array}{cc}
\mu-k^{2}+\gamma(k)\cos(\chi)-i(n+\zeta)\nu & -\sigma+\beta k^{2}\\
\sigma-\beta k^{2} & \mu-k^{2}-\gamma(k)\cos(\chi)-i(n+\zeta)\nu\end{array}\right)\left(\begin{array}{c}
X_{n}\\
Y_{n}\end{array}\right)\right.\label{eq:inflin}\\
 & \left.+\frac{\gamma(k)\sin(\chi)}{2}\left[\left(\begin{array}{cc}
1 & i\\
i & 1\end{array}\right)\left(\begin{array}{c}
X_{n+1}\\
Y_{n+1}\end{array}\right)+\left(\begin{array}{cc}
1 & -i\\
-i & 1\end{array}\right)\left(\begin{array}{c}
X_{n-1}\\
Y_{n-1}\end{array}\right)\right]\right\} =0,\nonumber \end{align}
 which can be written as a matrix equation: \begin{equation}
{\bf {\bf M(\zeta)v}}=\gamma(k){\bf Lv},\label{eq:GenEigEq}\end{equation}
 where \[
{\bf v}=\left(...\,\, X_{n-1}\,\, Y_{n-1}\,\, X_{n}\,\, Y_{n}\,\, X_{n+1}\,\, Y_{n+1}\,\,...\right)^{T}.\]
Here ${\bf M}$ is block-diagonal and ${\bf L}$ couples adjacent
modes $(X_{n},Y_{n})$ and $(X_{n\pm1},Y_{n\pm1})$. Equation (\ref{eq:GenEigEq})
is a generalized eigenvalue problem for the eigenvalue $\gamma(k)$,
which represents the forcing amplitude at the onset of the instability
for a given $k$. In our formulation of the linear problem $\gamma$
is assumed to be real; complex eigenvalues therefore do not correspond
to solutions of the original problem. To solve for the eigenvalues
for a given value of the wave number $k$, we truncate the sum in
(\ref{eq:GenEigEq}) at some $N$ and calculate the eigenvalues. We
test for convergence by requiring that for the real eigenvalues $\gamma_{i}^{(N)}$
the total change in all eigenvalues when $N$ is increased by $1$
is smaller than a tolerance $\Delta$, $\sum(\gamma_{i}^{(N+1)}-\gamma_{i}^{(N)})^{2}\leq\Delta=10^{-10}$.
We find that generally $N=10$ is sufficient. Repeating this process
over a range of $k$ we construct the neutral stability curves $\gamma^{(H)}(k)$
for the harmonic mode ($\zeta=0$) and $\gamma^{(S)}(k)$ for the
subharmonic mode ($\zeta=1/2$). The global minimum of these
curves yields the respective critical values $(k_{c}^{(H)},\gamma_{c}^{(H)})$
and $(k_{c}^{(S)},\gamma_{c}^{(S)})$.

\begin{figure}[h!]
\begin{centering}
(a) \includegraphics[width=2.5in]{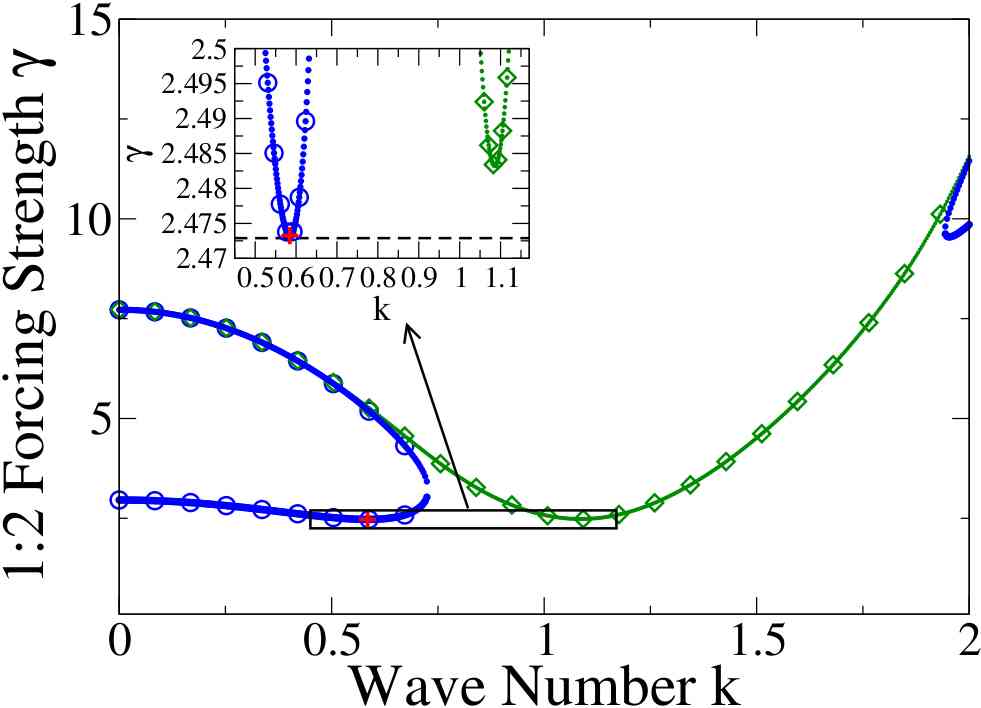}
\,\,\,\,\,\,\,\,(b) \includegraphics[width=2.5in]{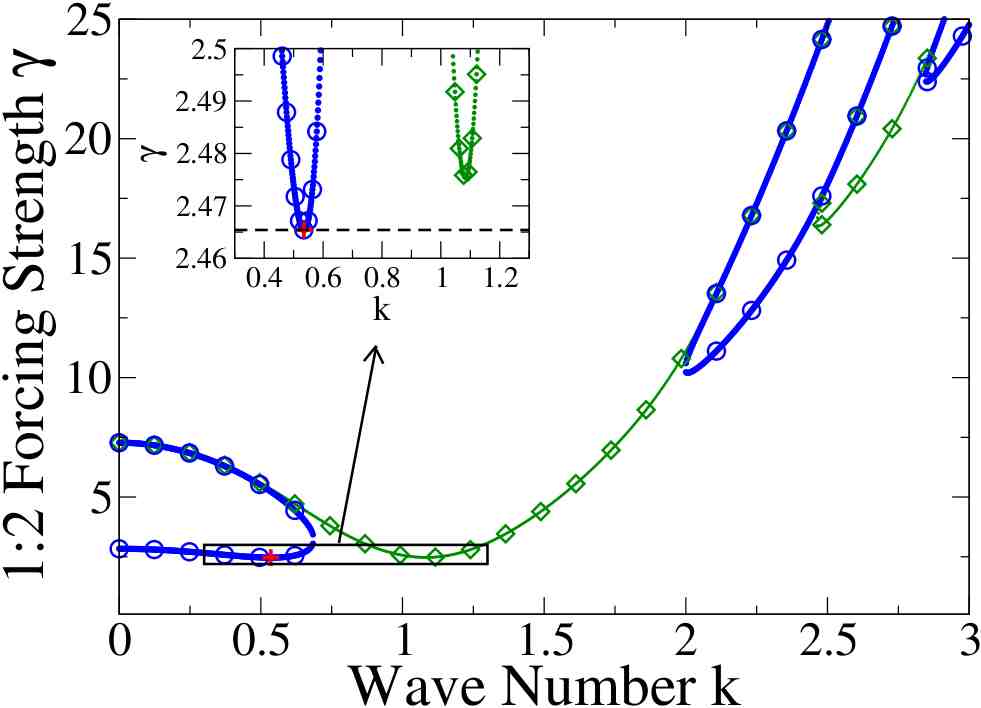}
\par\end{centering}

\caption{\label{neutral1}Harmonic (green, diamonds) and subharmonic (blue,
circles) neutral stability curves for $\mu=-1$, $\sigma=4$, $\beta=3$.
The parameters $\chi$ and $\nu$ are tuned to obtain the desired
$K$. (a) For $K=1.857$, $\chi=0.4883189$ and $\nu=3.824884$. (b)
For $K=2$,  $\chi=0.4767180$ and $\nu=4.2$.}
\end{figure}

The weakly nonlinear analysis below shows that weakly damped harmonic
modes have a strong impact on the selection of subharmonic patterns
via resonant triad interactions. Our aim is to exploit this sensitivity
to stabilize complex patterns like superlattice patterns. We therefore
focus here on parameters for which the minimum of the harmonic mode
is only slightly above that of the subharmonic mode;  we use $\gamma_{c}^{(H)}-\gamma_{c}^{(S)}=0.01$.
The resonant triads rely on spatiotemporal resonances that depend
decisively on the wavenumber ratio $K\equiv k_{c}^{(H)}/k_{c}^{(S)}$.
We consider the two cases $K=2\cos(\tan^{-1}(2/5))=1.857$ and $K=2$
and show in Fig.\ref{neutral1} the corresponding neutral stability
curves for $\gamma_{c}^{(H)}-\gamma_{c}^{(S)}=0.01$.

\begin{figure}[h!]
\begin{centering}
(a) \includegraphics[width=2.5in]{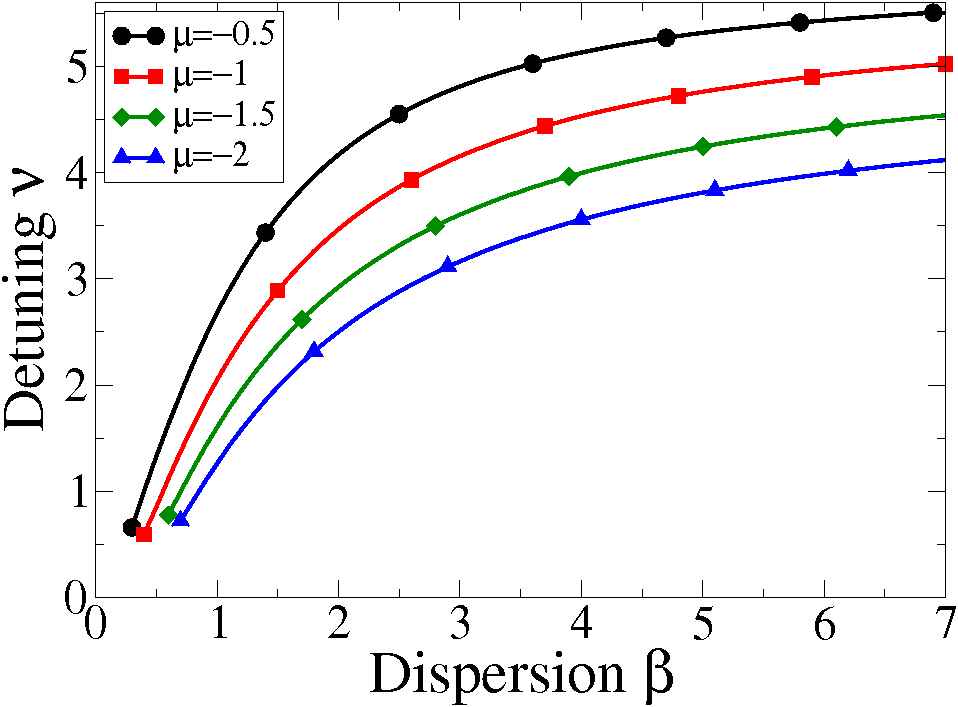}\,\,\,\,\,\,\,\,(b)
\includegraphics[width=2.5in]{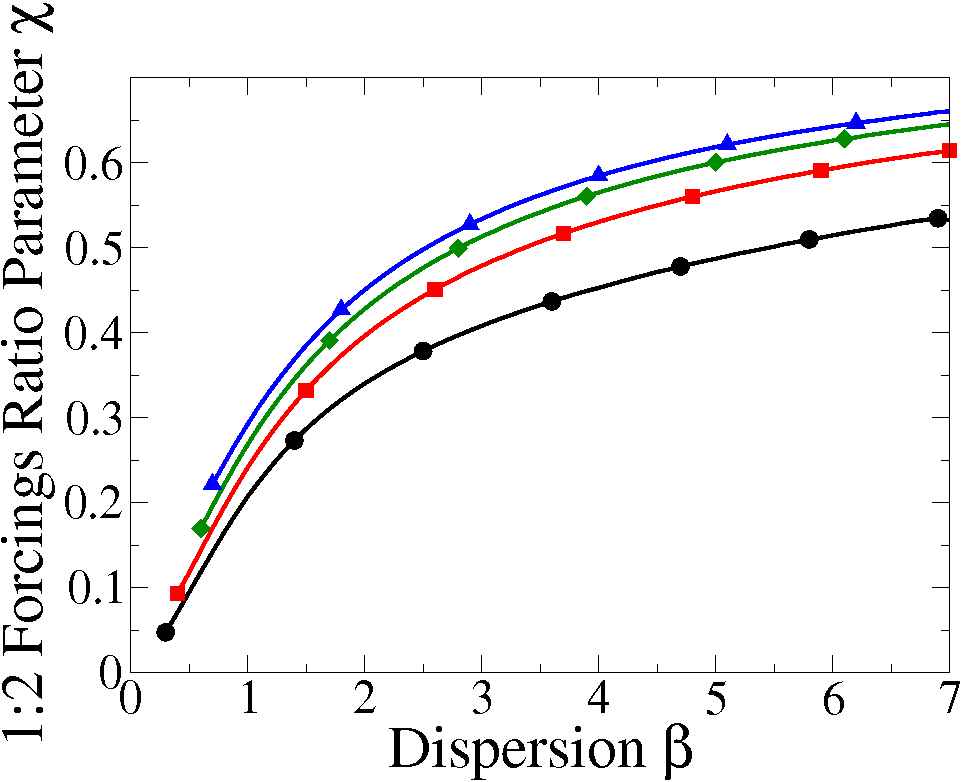}\\
 (c) \includegraphics[width=2.5in]{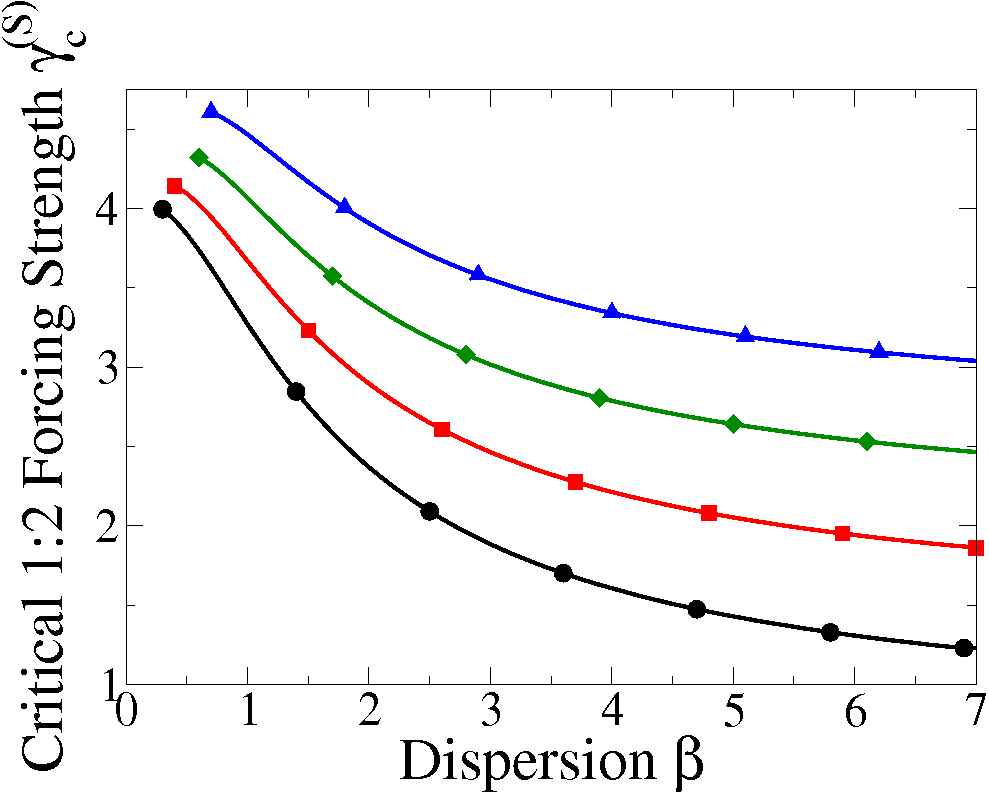}\,\,\,\,\,\,\,\,(d)
\includegraphics[width=2.5in]{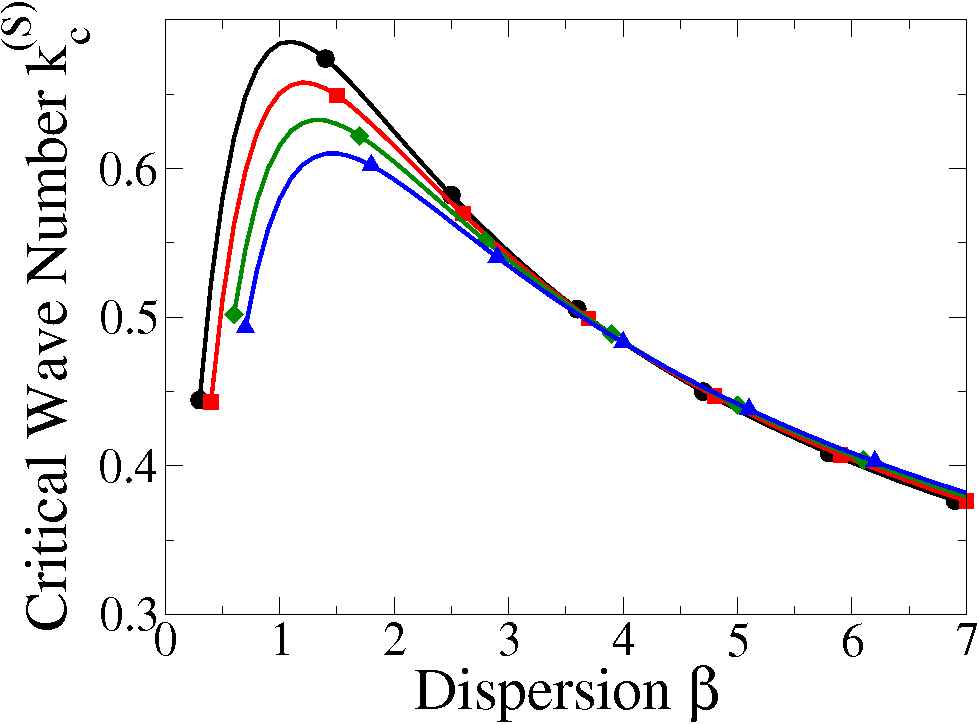}
\par\end{centering}

\caption{Linear parameters at the codimension-2 point $K=2$ and $\gamma_{c}^{(H)}-\gamma_{0}^{(S)}\equiv0.01$
for detuning $\sigma=4$ as a function of the dispersion $\beta$
and the damping $\mu$. (a) Detuning $\nu$, (b) forcing ratio $\chi$,
(c) critical forcing strength $\gamma_{c}^{(S)}$, and (d)  critical
wave number $k_{c}^{(S)}$   \label{Fig:LinRes_kratio2}}
\end{figure}

Varying the parameter $\chi$ characterizing the ratio of forcing
strengths near the 1:2-resonance shifts the critical forcing parameters
$\gamma_{c}^{(H)}$ and $\gamma_{c}^{(S)}$ relative to each other,
while varying the parameter $\nu$ characterizing the detuning difference
shifts the critical wavenumbers $k_{c}^{(H)}$ and $k_{c}^{(S)}$.
Requiring $\gamma_{c}^{(H)}-\gamma_{0}^{(S)}=0.01$ for a fixed value
of $K$ defines a codimension-2 point that determines both $\chi$
and $\nu$. For fixed $K=2$ their dependence on the dispersion $\beta$
and the damping $\mu$ is illustrated in Fig.\ref{Fig:LinRes_kratio2}
along with the corresponding values of $\gamma_{c}^{(S)}$ and $k_{c}^{(S)}$.
Since $\bar{A}$ satisfies the  CGLE with the opposite signs of the
imaginary parts of the equation, only positive values of $\beta$
need to be considered.

For $\chi=0$ the critical wave number is given by $k_{c}^{2}=(\mu+\nu\beta)/(1+\beta^{2})$
corresponding to a critical forcing $\gamma_{c}^{2}=(\nu-\mu\beta)^{2}/(1+\beta^{2})$
\cite{CoFr94}.  Thus, $k_{c}$ has a local maximum and $k_{c}^{2}\sim1/\beta$
for large $\beta$; qualitatively the same behaviour is found at the
codimension-2 point, as illustrated in Fig.\ref{Fig:LinRes_kratio2}d.
Fig.\ref{Fig:LinRes_kratio2}c confirms that the amount of forcing
necessary to generate patterns increases as one goes further below
the Hopf bifurcation, i.e. as $\mu$ becomes more negative. 

\section{Weakly Nonlinear Analysis}

\label{sec:Weakly-Nonlinear-Analysis}

To determine the stability of the desired subharmonic patterns we
derive the corresponding amplitude equations by performing a weakly
nonlinear analysis of (\ref{eq:cgle_final}). We then use energy arguments
to guide us in terms of the relative stability of the various pattern
comprised of different numbers of modes. We are in particular interested
in the stability of patterns comprising three or more modes.

\subsection{Amplitude Equations}

\label{sub:Amplitude-Equations}

Splitting the CGLE (\ref{eq:cgle_final}) again into real and imaginary
parts, we obtain a system of real partial differential equations.
For a subharmonic pattern with $N$ modes, we expand $(A_{r},A_{i})$
about $(0,0)$ as\begin{align}
\left(\begin{array}{c}
A_{r}\\
A_{i}\end{array}\right)= & \epsilon\left\{ \sum_{m=1}^{N}Z_{m}(T)\left(\sum_{n=-\infty}^{\infty}\left(\begin{array}{c}
X_{n}\\
Y_{n}\end{array}\right)e^{i(n+\frac{1}{2})\nu t}\right)e^{i{\bf k}_{m}\cdot{\bf x}}+c.c.\right\} \nonumber \\
 & +\epsilon^{2}\left(\begin{array}{c}
A_{r}^{(2)}\\
A_{i}^{(2)}\end{array}\right)+\epsilon^{3}\left(\begin{array}{c}
A_{r}^{(3)}\\
A_{i}^{(3)}\end{array}\right)+\dots,\,\,\,|{\bf k}_{m}|=k_{c}\label{eq:WNLansatz}\end{align}
 where $0<\epsilon\ll1$ and the complex amplitudes $Z_{m}(T)$ depend
on the slow time $T=\epsilon^{2}t$. The corresponding wave vectors
are denoted by ${\bf k}_{m}$. We also expand $\gamma$ as $\gamma=\gamma_{c}+\epsilon^{2}\gamma_{2}$.

A standard but lengthy calculation  yields amplitude equations for
the $Z_{j}(T)$ describing the $N-$mode pattern:

\begin{equation}
\frac{dZ_{i}}{dT}=\lambda\gamma_{2}Z_{i}-\left(b_{0}|Z_{i}|^{2}+\sum_{j=1,j\neq i}^{N}b(\theta_{ij})|Z_{j}|^{2}\right)Z_{i},\,\,\, i=1\dots N,\label{eq:GenAmpEqs}\end{equation}
 where $\theta_{ij}$ corresponds to the angle between the wave vectors
${\bf k}_{i}$ and ${\bf k}_{j}$. Note that there are no quadratic
terms in (\ref{eq:GenAmpEqs}),  because these amplitude equations
correspond to subharmonic patterns and must therefore be equivariant
under the symmetry $Z_{j}\rightarrow-Z_{j}$. In order to have stable
complex patterns made up of many modes, the competition between modes
needs to be sufficiently weak.  Since in the amplitude equations (\ref{eq:GenAmpEqs})
only pairwise mode interactions arise, a minimal condition for complex
patterns to be stable is the stability of  rectangle patterns, which
are comprised of $2$ modes at some angle $\theta$, with respect
to stripes. This requires $|b(\theta)/b_{0}|<1$.

\begin{figure}[h!]
\begin{centering}
(a) \includegraphics[width=2.5in]{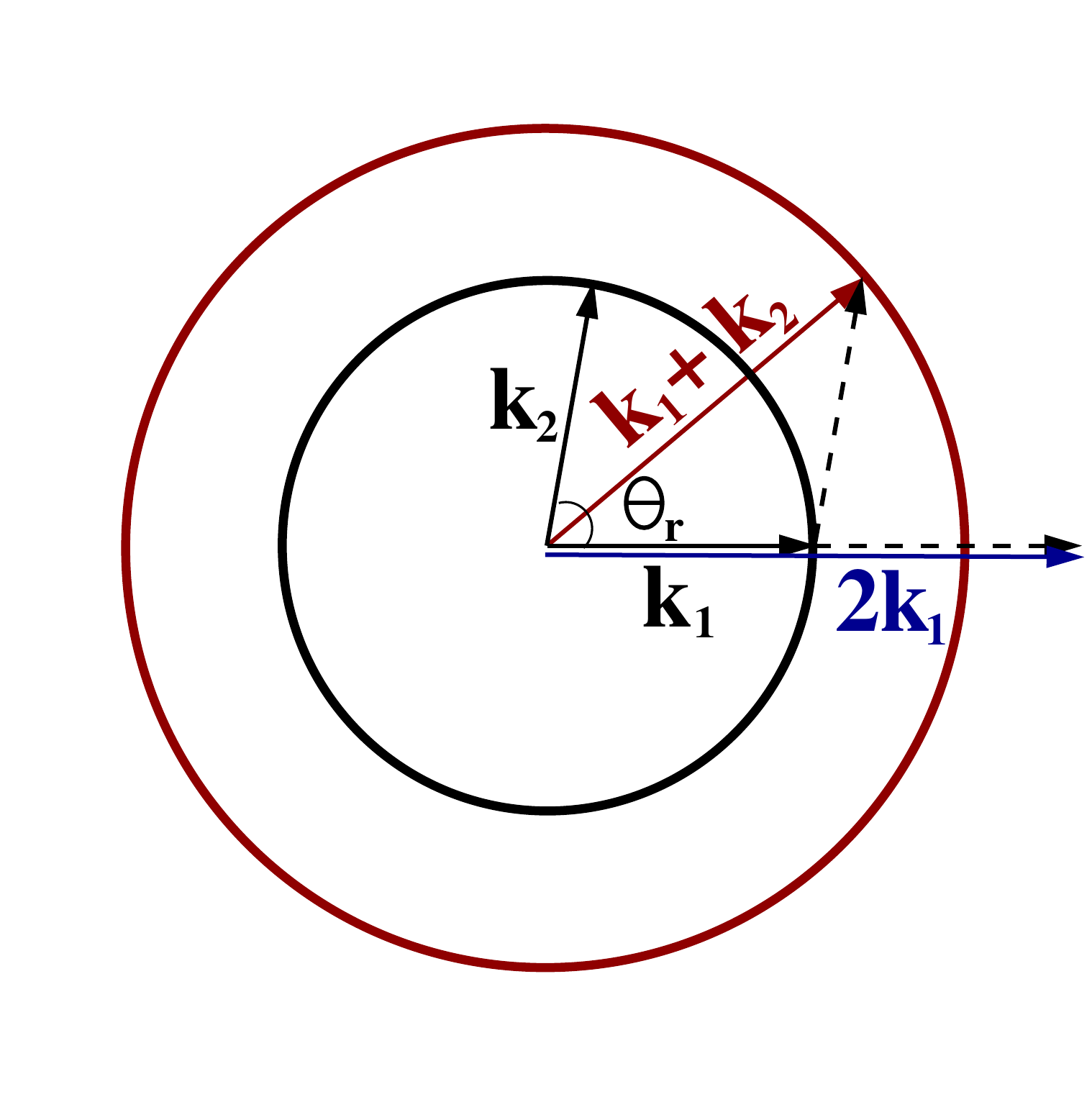}\,\,\,\,\,\,\,\,
(b) \includegraphics[width=2.5in]{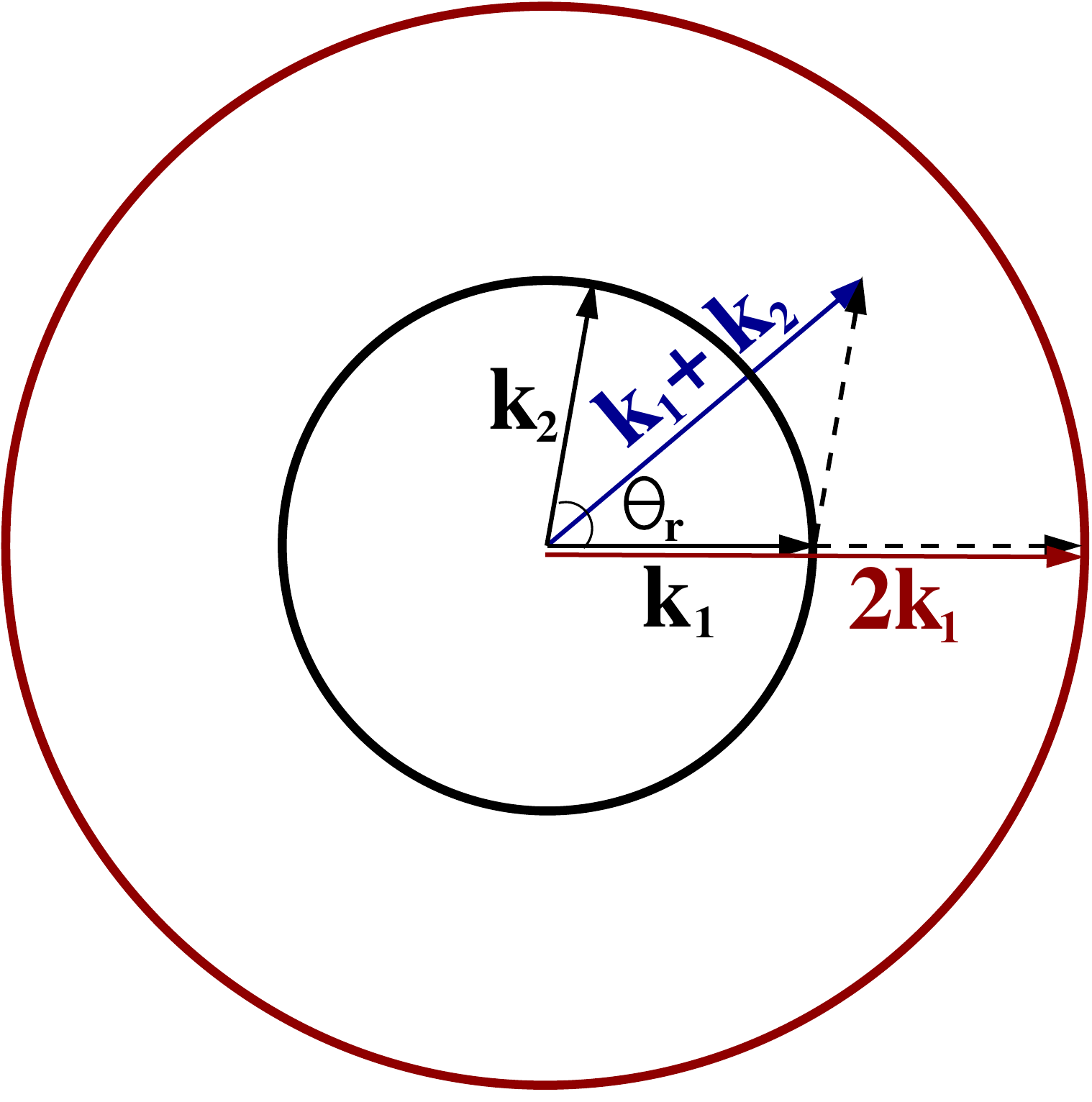}
\par\end{centering}

\caption{Resonant triad interaction. Inner circle represents critical circle
at $k_{c}^{(S)},$ outer circle represents weakly damped mode at (a)
$K=2\tan^{-2}(2/5)$, (b) $K=2$. The wave vector ${\bf \textbf{k}}_{1}+{\bf \textbf{k}}_{2}$
is relevant for $b(\theta)$ while $2{\bf \textbf{k}}_{1}$ affects
$b_{0}$ (cf. eq.(\ref{eq:GenAmpEqs})). \label{Fig:ResTriad}}
\end{figure}

The self- and cross-coupling coefficients, $b_{0}$ and $b(\theta_{ij})$,
respectively, can be strongly influenced by resonant triad interactions
as illustrated in Fig.\ref{Fig:ResTriad}. There the black circles
represent the subharmonic critical circle and the red circle the wavevector
of the most weakly damped harmonic mode for a given choice of critical
forcing $\gamma$. At quadratic order resonant triad interaction takes
place through two mechanisms, each feeding into the coefficients at
cubic order. The first mechanism is through the interaction of two
different wave vectors (${\bf \textbf{k}}_{1}+{\bf \textbf{k}}_{2}$
in Fig.\ref{Fig:ResTriad}a) seperated by an angle $\theta_{r}$:
the cross-coupling coefficient $b(\theta)$ is strongly impacted near
$\theta=\theta_{r}$ if the mode excited at quadratic order through
the interaction of these two wave vectors is weakly damped \cite{SiTo00,PoSi03,PoTo04,RuSi07}.
The angle $\theta_{r}$ is a function of the wavenumber ratio $K$,
$\theta_{r}=2\cos^{-1}(K/2)$. The second mechanism is through
the interaction of a wave vector with itself ($2\mathbf{k}_{1}$ in
Fig.\ref{Fig:ResTriad}b): in this case, if the harmonic mode at $2\mathbf{k}_{c}$
is weakly damped,  the self-coupling coefficient $b_{0}$ is strongly
influenced \cite{ZhVi96,ZhVi97,RuSi07}. In Fig.\ref{Fig:ResTriad}a
$K<2$ ($\theta_{r}\neq0$), so the mode excited by the vector ${\bf \textbf{k}}_{1}+{\bf \textbf{k}}_{2}$
is weakly damped and the mode excited by the $2{\bf \textbf{k}}_{1}$
vector is strongly damped. In Fig.\ref{Fig:ResTriad}b $K=2$ ($\theta_{r}=0$),
so the opposite is true.

\begin{figure}[h!]
\begin{centering}
(a) \includegraphics[width=2.5in]{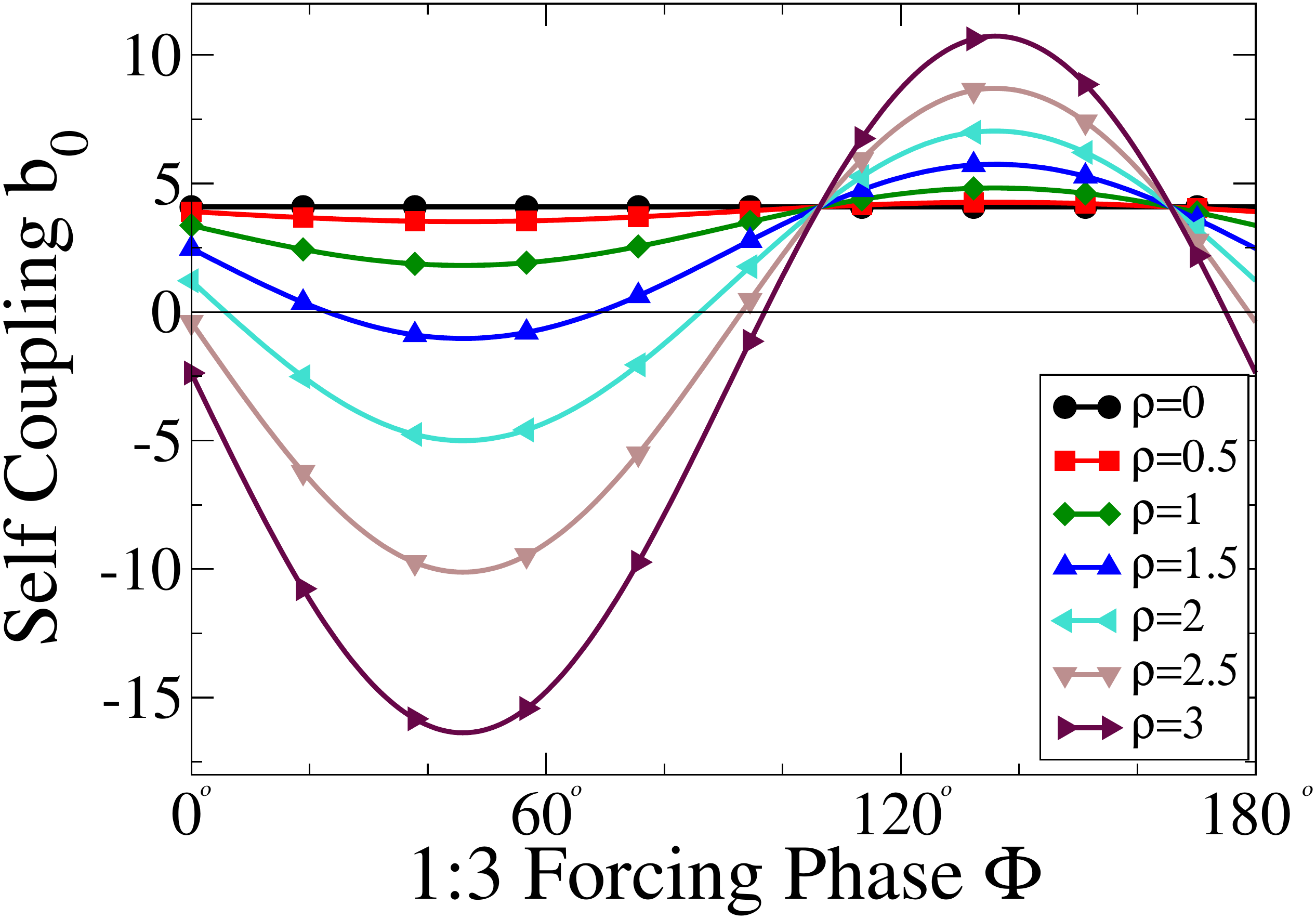}\,\,\,\,\,\,\,\,(b)
\includegraphics[width=2.5in]{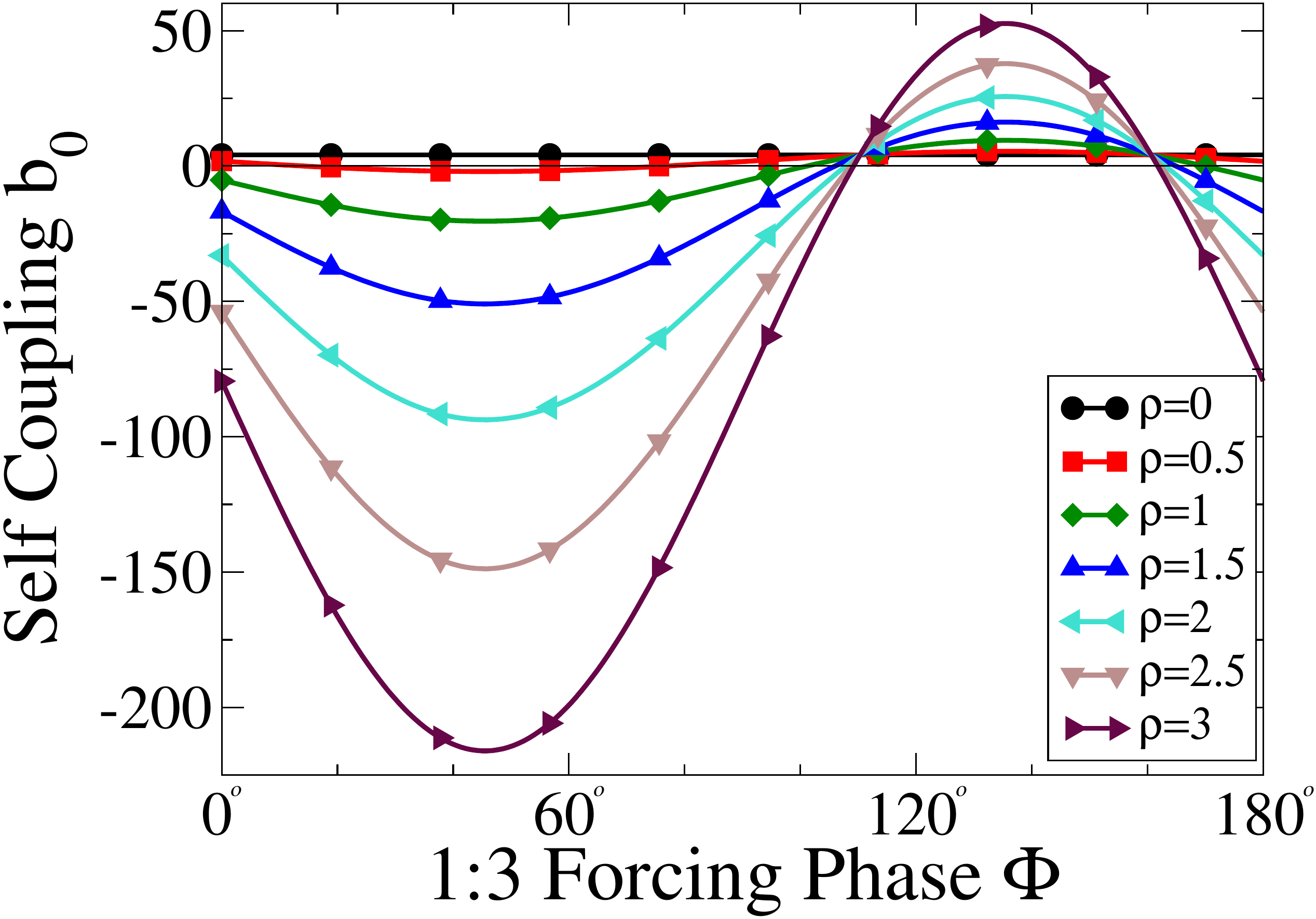} 
\par\end{centering}

\caption{Self-coupling coefficient $b_{0}$ plotted against the 1:3 forcing
phase $\Phi$ for various values of the 1:3-forcing strength $\rho$
with $\alpha=-1$. (a)  $K=2\cos(\tan^{-1}(2/5)$, other parameters
as in Fig.\ref{neutral1}a; (b)  $K=2$, other parameters as in Fig.\ref{neutral1}b.\label{Fig:SelfCoupling}}
\end{figure}

Fig.\ref{Fig:SelfCoupling} shows the effect of the resonant triad
interactions on the self-coupling coefficient $b_{0}$, by plotting
\textbf{$b_{0}$} as a function of the 1:3 forcing phase $\Phi$ for
different values of the 1:3 forcing strength $\rho$. Fig.\ref{Fig:SelfCoupling}a
corresponds to the case illustrated in Fig.\ref{Fig:ResTriad}a with
$K=2\cos(\tan^{-1}(2/5))$; with this choice of $K$ the resonance
angle $\theta_{r}\equiv2\tan^{-1}(2/5)$ is near $\pi/4$ and the
associated Fourier modes fall on a regular grid \cite{RuRu03} for
later comparison with numerical simulation. In Fig.\ref{Fig:SelfCoupling}b
$K=2$ and the effect of the resonant triad on $b_{0}$ is greatly
enhanced - note the difference in scale compared to Fig.\ref{Fig:SelfCoupling}a
- and even quite weak forcing $\rho$ can change the sign of $b_{0}$
and with it the direction of the bifurcation to stripes. In the following
we  restrict ourselves to values of the phase $\Phi$ for which  $b_{0}>0$
and the stripes bifurcate supercritically. 

\begin{figure}[h!]
\begin{centering}
\includegraphics[width=2.5in]{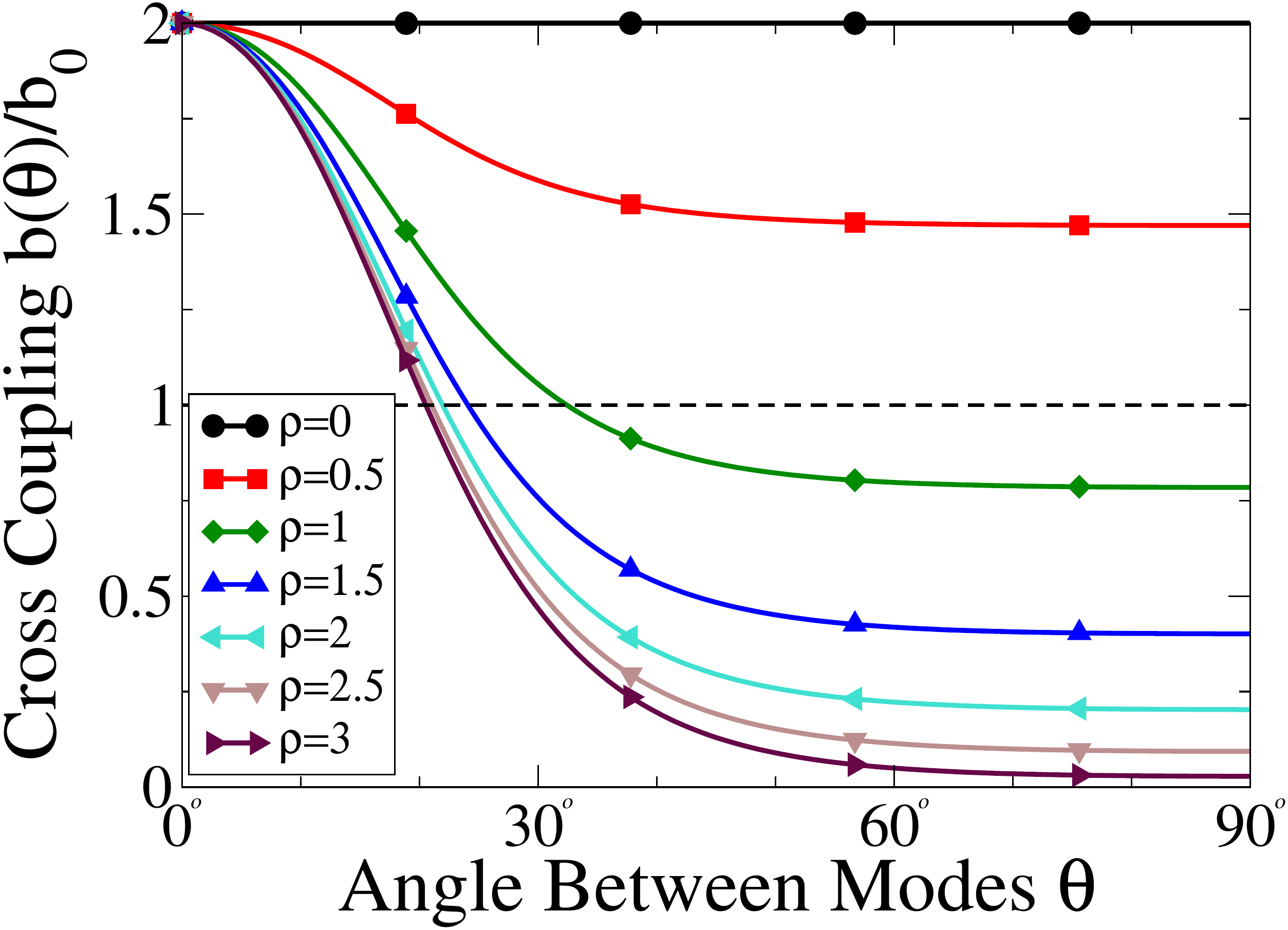}
\par\end{centering}

\caption{Coupling coefficient ratio $b(\theta)/b_{0}$ for $K=2$ with $\Phi=3\pi/4$
with linear parameters as in Fig.\ref{neutral1}b and nonlinear dispersion
$\alpha=-1$ for different strengths of 1:3-forcing $\rho$. Rectangles
are stable for $|b(\theta)/b_{0}|<1$. \label{Fig:Cubic Coefs Kis2} }
\end{figure}

To assess the mode competition for $K=2$, Fig.\ref{Fig:Cubic Coefs Kis2}
shows the dependence of the cross-coupling coefficient ratio $b(\theta)/b_{0}$
on $\theta$.  To enhance the stability of patterns comprised of multiple
modes by maximizing the range of weak competition we maximize the
damping $b_{0}$ and  choose $\Phi=3\pi/4$ (cf.  Fig.\ref{Fig:SelfCoupling}).
 Since $K=2$, the self-coupling $b_{0}$ is enhanced by the resonant
triad, while away from $\theta=\theta_{r}\equiv0$ the cross-coupling
$b(\theta)$ is only weakly affected. Thus, as expected, for $\rho\ge1$
the ratio $b(\theta)/b_{0}$ is strongly reduced away from $\theta=0$
allowing for rectangle patterns corresponding to angles as small as
$\theta=\theta_{c}\sim30^{\circ}$ and smaller to be stable with respect
to stripes. Consequently, patterns comprised of four or possibly even
more modes are expected to be stable, as discussed in Sec.\ref{sub:Energy}.

The mode interaction also depends on the linear dispersion $\alpha$
and the nonlinear dispersion parameter $\beta$. These two parameters
are system parameters that may not be tunable in experiments. It is
therefore of interest to assess over what range in these parameters
complex patterns are to be expected and to what extent the forcing
parameters can be adjusted so as to reach the regions of interest
for given system parameters. 

\begin{figure}[h!]
(a) \includegraphics[width=2.5in]{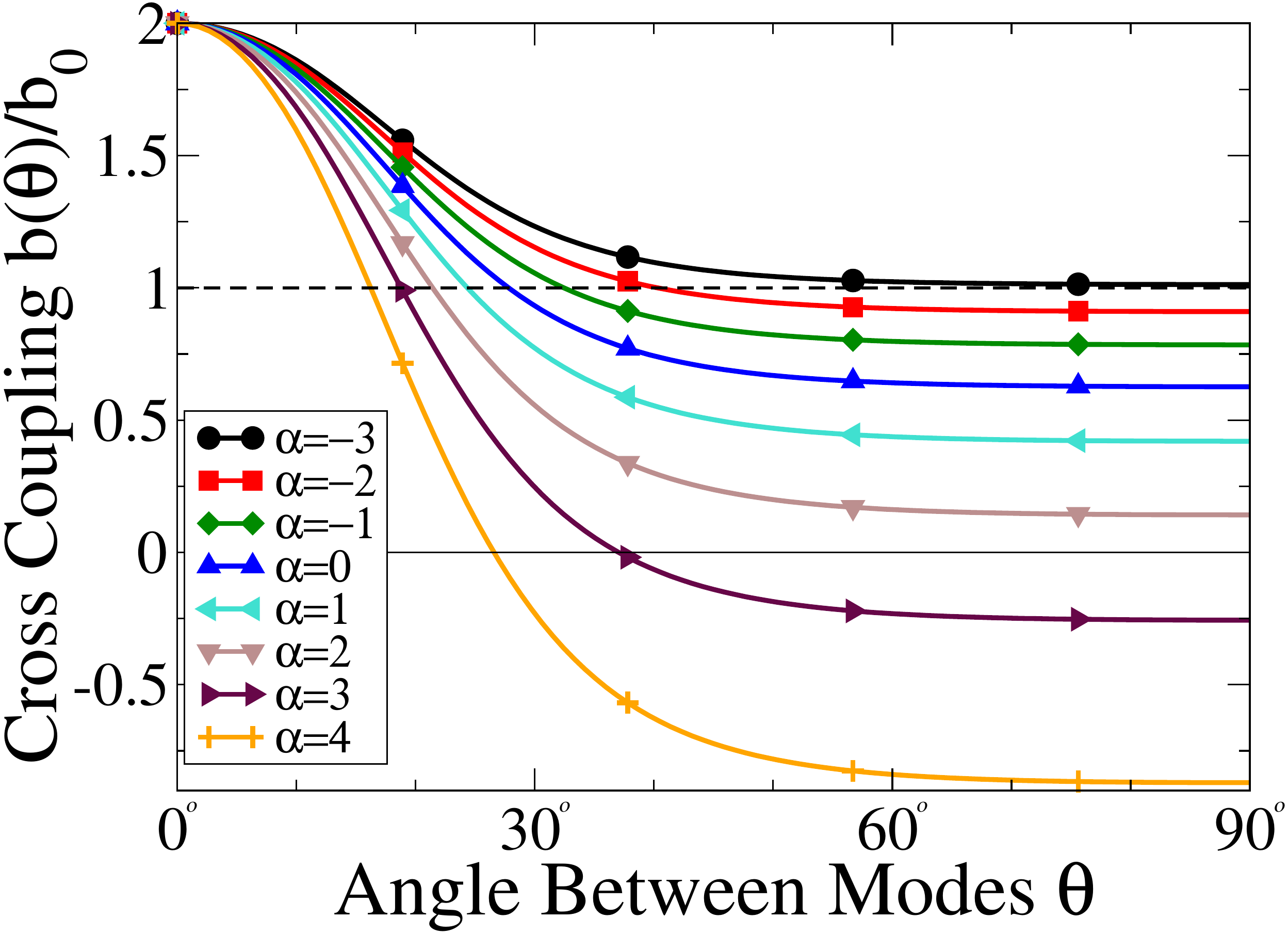}\,\,\,\,\,\,\,\,(b) \includegraphics[width=2.5in]{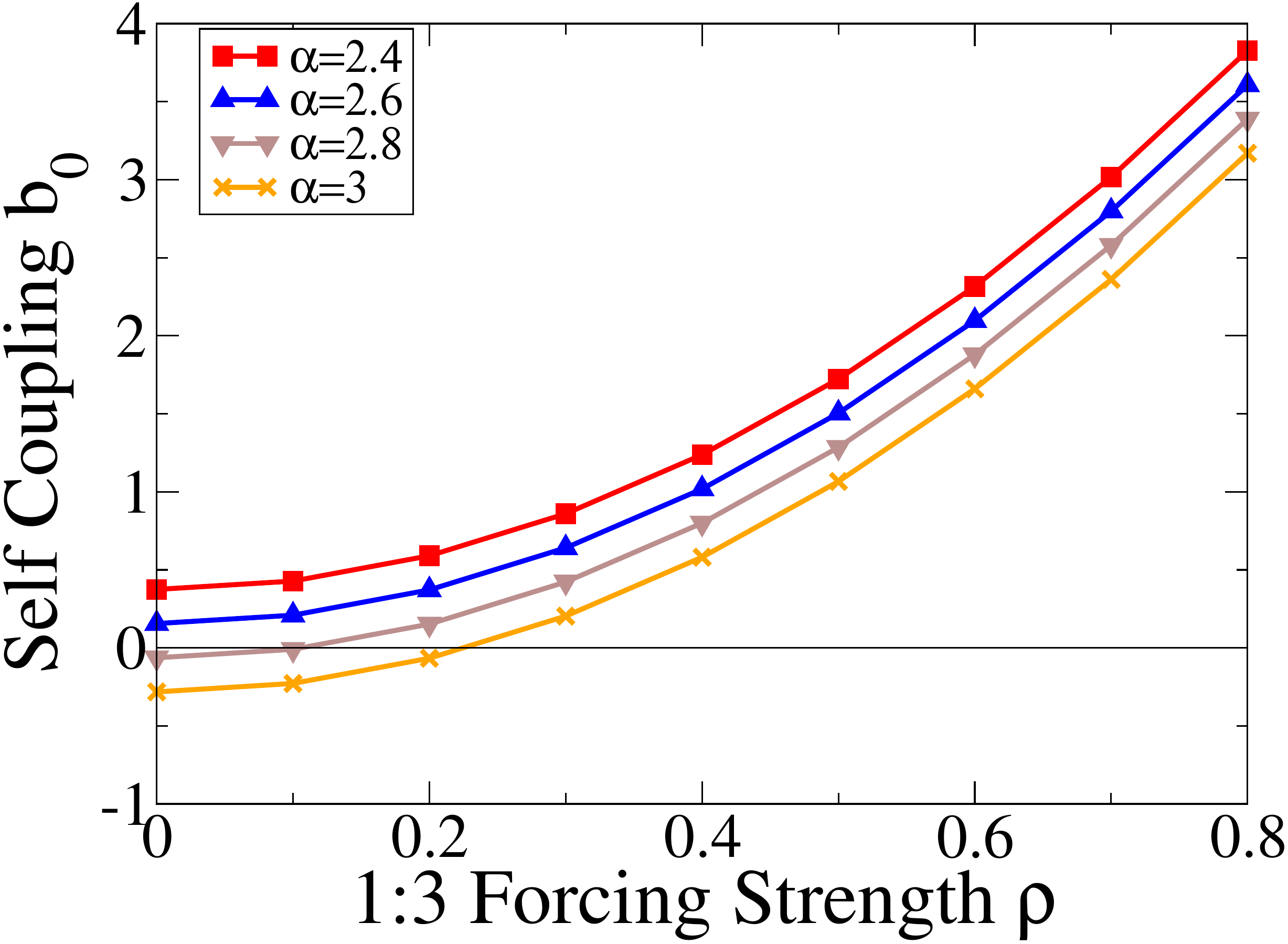}

\caption{Coupling coefficient ratio $b(\theta)/b_{0}$ (a) and self-coupling
$b_{0}$ (b) for $K=2$ and $\Phi=3\pi/4$ with parameters as in Fig.\ref{neutral1}b
for 1:3 forcing strength $\rho=1$ and different values of the nonlinear
dispersion parameter $\alpha$. \label{fig:alpha-dependence}}
\end{figure}

The effect of the nonlinear dispersion parameter $\alpha$ on the
ratio $b(\theta)/b_{0}$ is shown in Fig.\ref{fig:alpha-dependence}a
for $\rho=1$. Somewhat similar to the dependence on $\rho$, with
increasing $\alpha$ the ratio $b(\theta)/b_{0}$ decreases and the
range of $\theta$ for which rectangle patterns are stable increases.
For more positive values of $\alpha$ even small forcing amplitudes
$\rho$ reduce the mode interaction significantly, but as $\rho$
is increased the ratio $b(\theta)/b_{0}$ saturates at values that
are not much smaller than shown in Fig.\ref{Fig:Cubic Coefs Kis2}
for $\alpha=-1$. In the simulations discussed in Sec.\ref{sec:Results}
below we use a moderate value of $\alpha=-1$. For single-frequency
forcing it was  found that the self-coupling coefficient $b_{0}$
changes sign for $\alpha=\beta$, rendering the bifurcation subcritical
for $\alpha>\beta$ \cite{CoFr94}. Interestingly, with multi-frequency
forcing the sign of $b_{0}$  depends also on the forcing strength
$\rho$ and, in fact, even for $\alpha=\beta$ the bifurcation to
stripes can be made supercritical by increasing the forcing strength.
This is illustrated in Fig.\ref{fig:alpha-dependence}b where $b_{0}$
is shown as a function of $\rho$ for different values of $\alpha$
near $\beta=3$.

\begin{figure}[h!]
\begin{centering}
\includegraphics[width=3in]{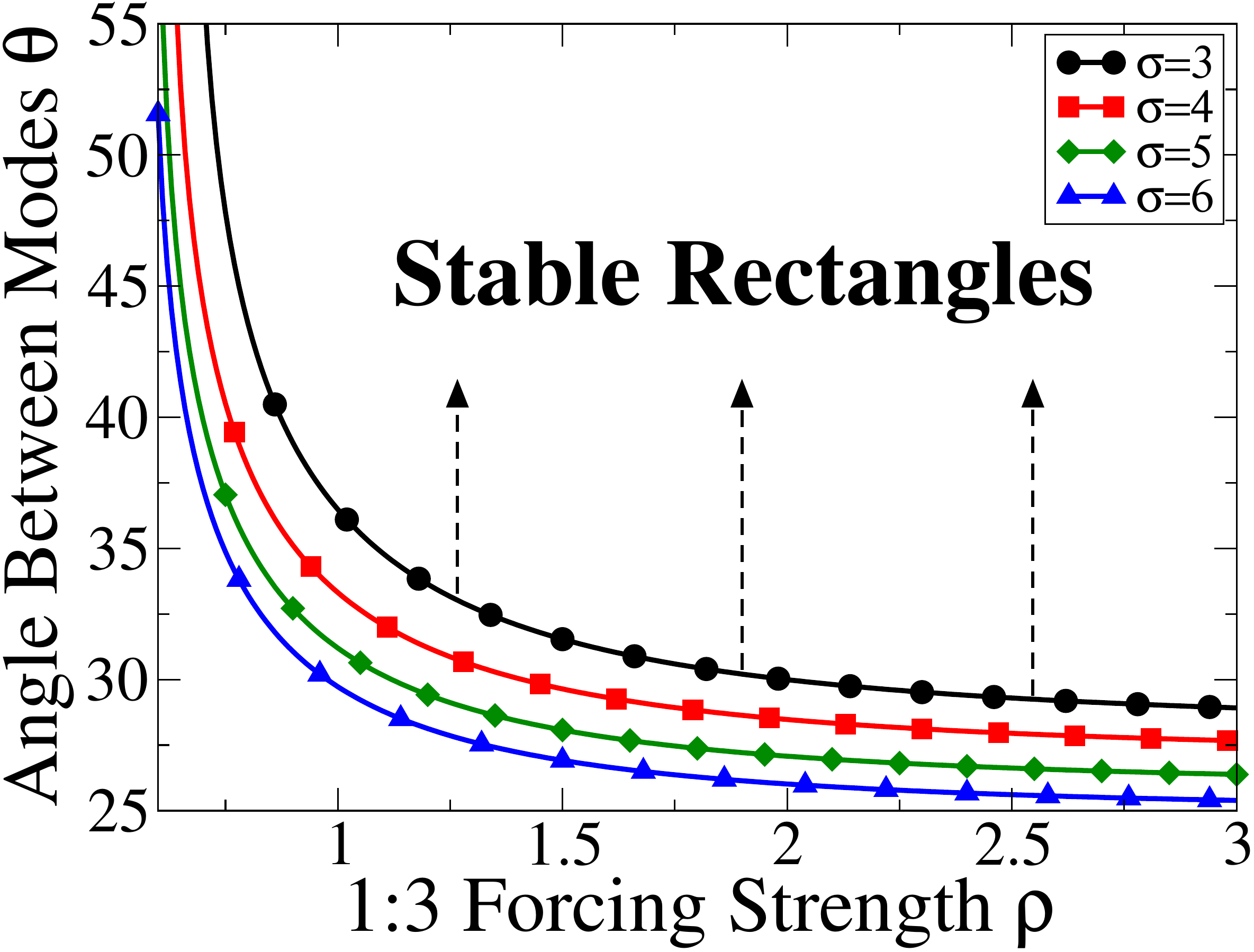} 
\par\end{centering}

\caption{Belousov-Zhabotinsky case. Minimal angle $\theta_{c}$ for which
rectangles are stable to stripes ($b(\theta_{c})/b_{0}=1$) as a function
of the 1:3-forcing strength $\rho$ for different values of the detuning
$\sigma$. Parameters $\mu=-1,$ $\alpha=-0.4$, $\beta=1.4$ with $\chi$
and $\nu$ tuned to the codimension-2 point $K=2$ and $\gamma_{c}^{(H)}-\gamma_{0}^{(S)}=0.01$.
.\label{fig:ThetaVsRho_CrossCouplingIs1}}
\end{figure}

To assess the dependence of the mode interaction on the linear dispersion
coefficient $\beta$ we do not perform a scan in $\beta$ but rather
focus on one other value, $\beta=1.4$, which corresponds to the value
found experimentally for the Belousov-Zhabotinsky reaction \cite{HySo93}.
We adjust the forcing parameters $\chi$ and $\nu$ to stay  at the
codimension-2 point $K=2$ and $\gamma_{c}^{(H)}-\gamma_{0}^{(S)}=0.01$.
The nonlinear dispersion coefficient was found in the experiments
to be \textbf{$\alpha=-0.4$}. Since the functional form of the angle
dependence of $b(\theta)$ does not change much  with the parameters
we use the minimal angle $\theta_{c}$ for which rectangles are still
stable with respect to stripes  as a proxy for the strength of the
mode competition. Thus $b(\theta_{c})/b_{0}=1$.  In Fig.\ref{fig:ThetaVsRho_CrossCouplingIs1}
we show $\theta_{c}$  as a function of the forcing strength $\rho$
for various values of the overall detuning $\sigma$. Clearly, even
for these experimentally relevant parameters the resonant triad interaction
can be exploited to make the mode competition sufficiently weak to
stabilize rectangle patterns over a wide range of angles.  This suggests
 that multi-mode patterns should become stable for moderate values
of the forcing parameters, which is confirmed below.

\begin{figure}[h!]
\begin{centering}
(a) \includegraphics[width=2.5in]{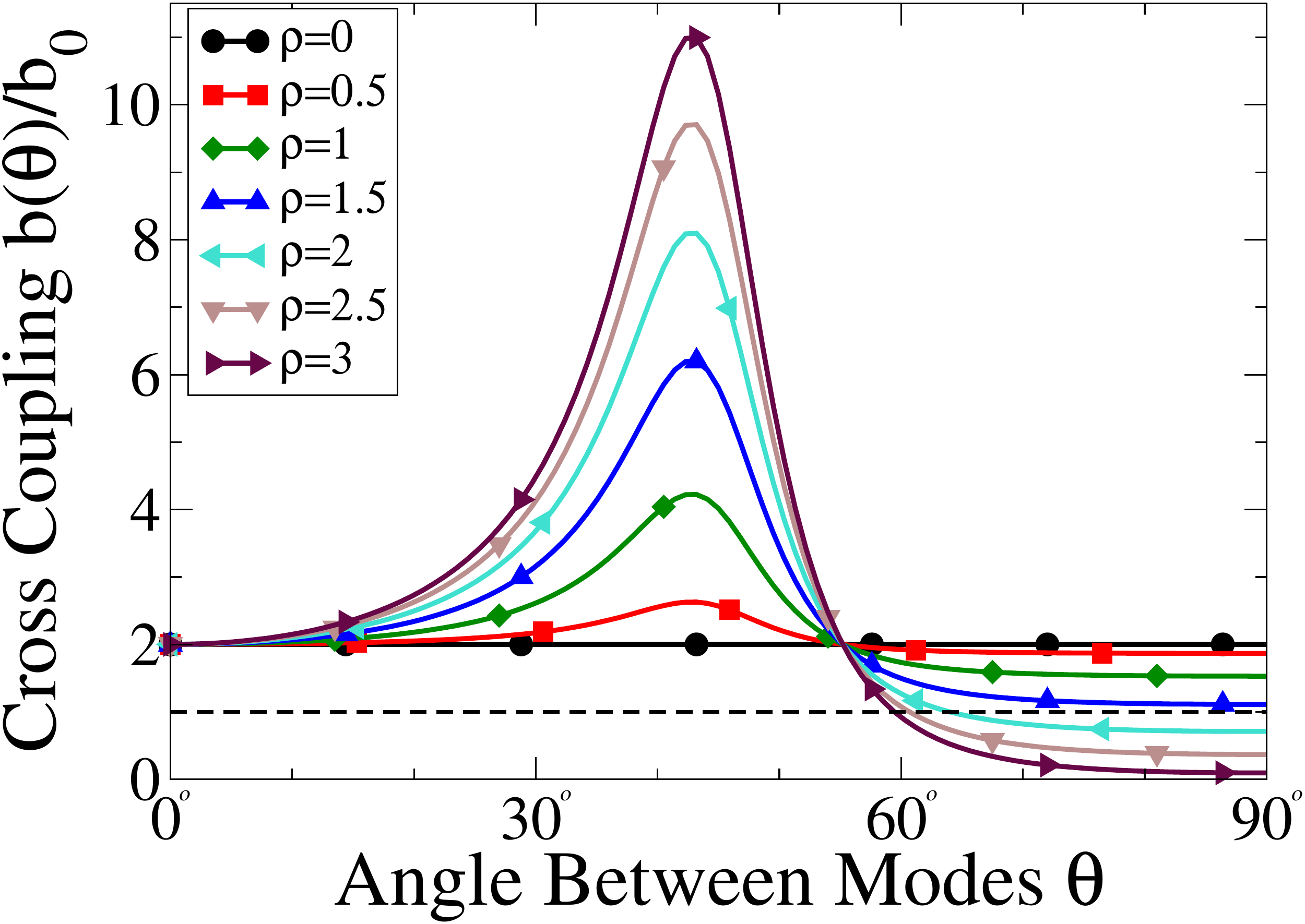}\,\,\,\,\,\,\,\,(b)
\includegraphics[width=2.5in]{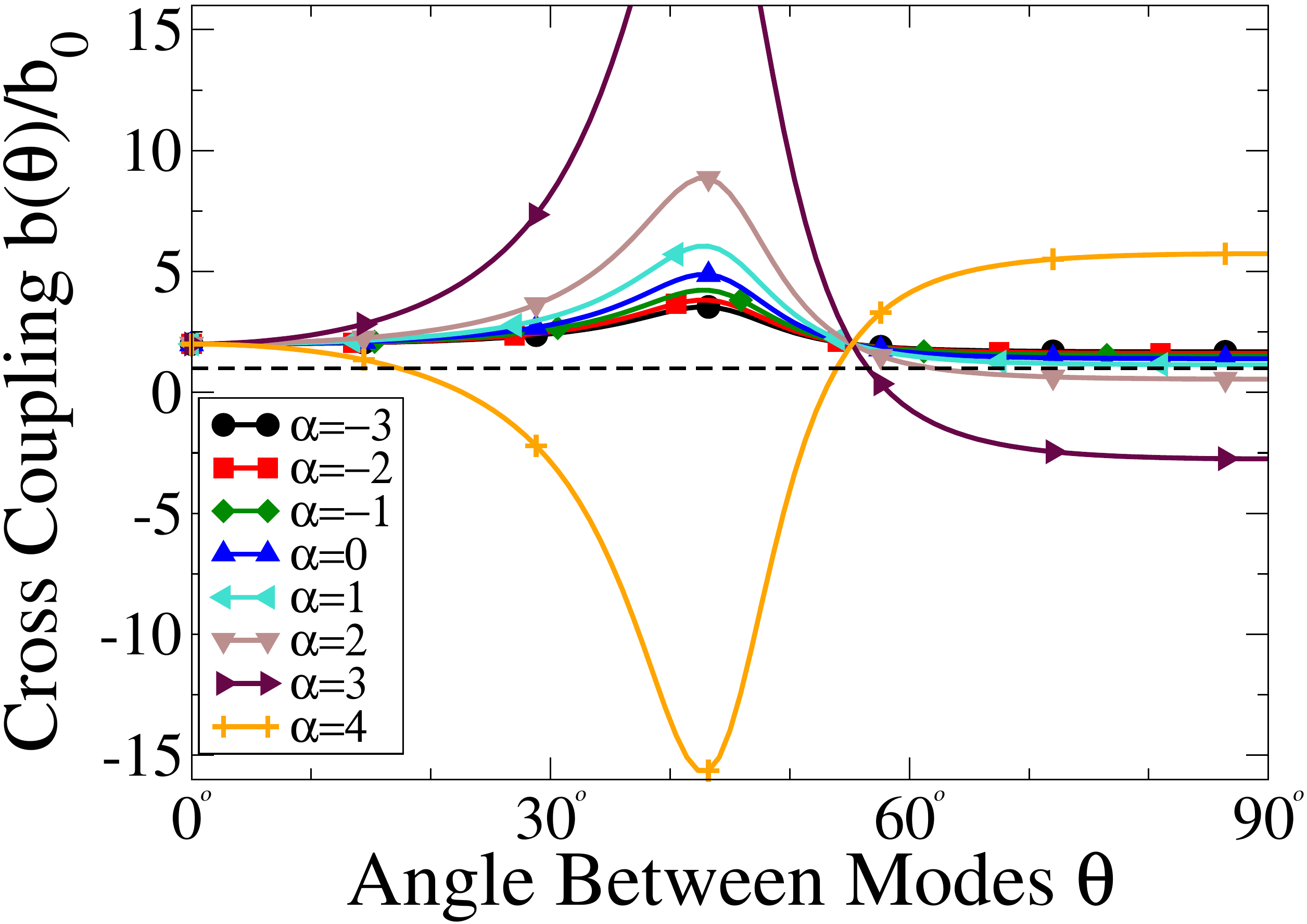}
\par\end{centering}

\caption{Coupling coefficient ratios  for $K=2\cos(\tan^{-1}(2/5))$
and $\Phi=3\pi/4$ with parameters as in Fig.\ref{neutral1}a for
(a) nonlinear dispersion parameter $\alpha=-1$ and different strengths
of 1:3-forcing $\rho$ and (b) 1:3 forcing strength $\rho=1$ and
different values of the nonlinear dispersion parameter $\alpha$.\label{fig:Cubic Coefs K is 1.857_phi3pio4}}
\end{figure}

For $K=2\cos(\tan^{-1}(2/5))$ the resonant triad affects
mostly $b(\theta)$ rather than $b_{0}$. Fig.\ref{fig:Cubic Coefs K is 1.857_phi3pio4}
shows the resulting ratio $b(\theta)/b_{0}$ as a function of the
angle $\theta$. To ensure $b_{0}>0$ we use again $\Phi=3\pi/4$.
As expected $b(\theta)/b_{0}$ exhibits a prominent peak near $\theta=\theta_{r}=2\tan^{-1}(2/5)$.
As in Fig.\ref{Fig:Cubic Coefs Kis2} the enhancement is increased
with $\rho$ (Fig.\ref{fig:Cubic Coefs K is 1.857_phi3pio4}a), and
as $\alpha$ becomes more positive (Fig.\ref{fig:Cubic Coefs K is 1.857_phi3pio4}b).
We also see from these plots that the range for which $b(\theta)/b_{0}<1$,
that is, the range of stable rectangles, is limited to  $60^{\circ}\lesssim\theta<90^{\circ}$.
Thus, we anticipate only stripe, square or hexagon patterns  to be
stable in this regime.

\begin{figure}[h!]
\begin{centering}
(a) \includegraphics[width=2.5in]{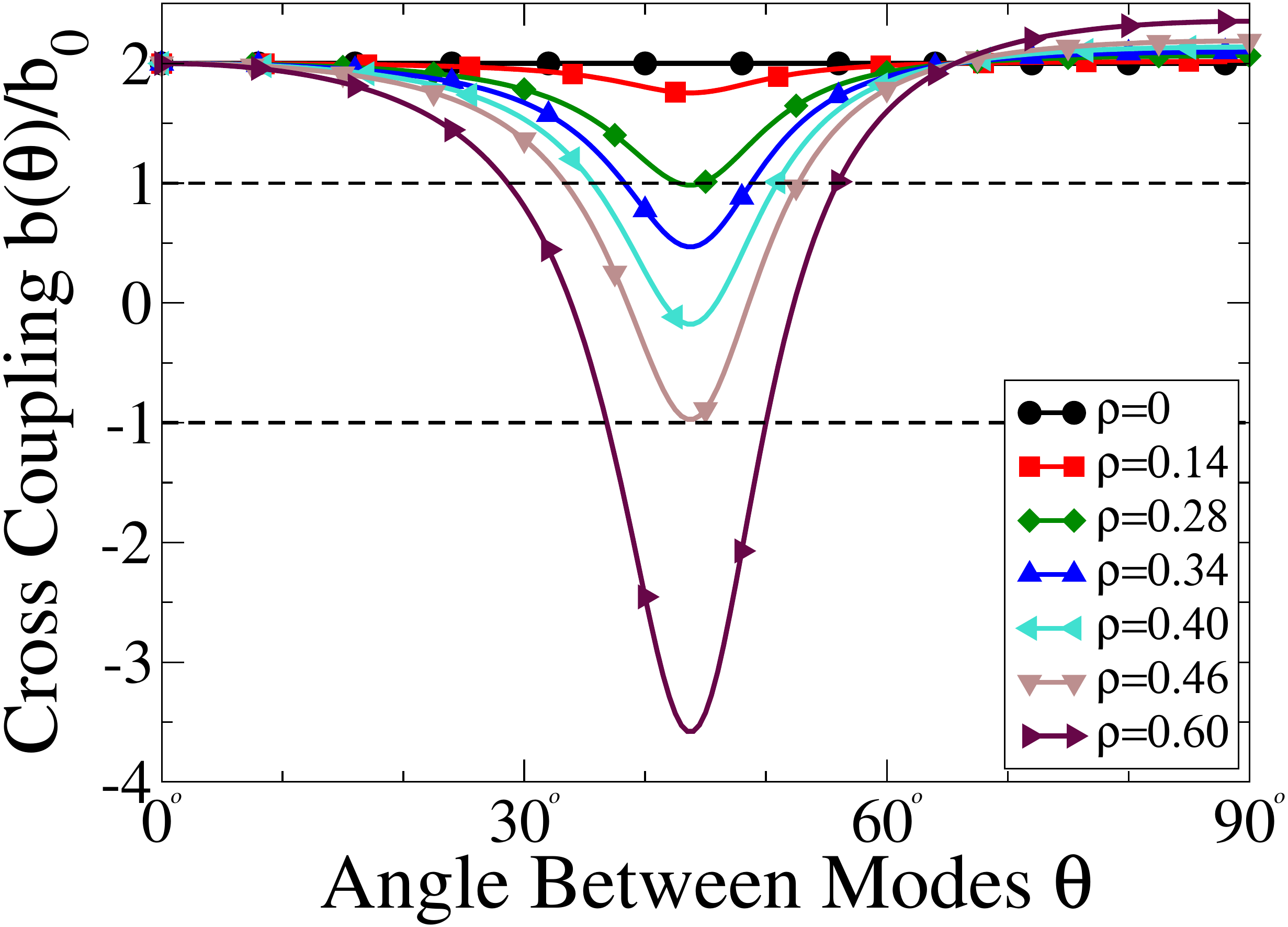}\,\,\,\,\,\,\,\,(b)
\includegraphics[width=2.5in]{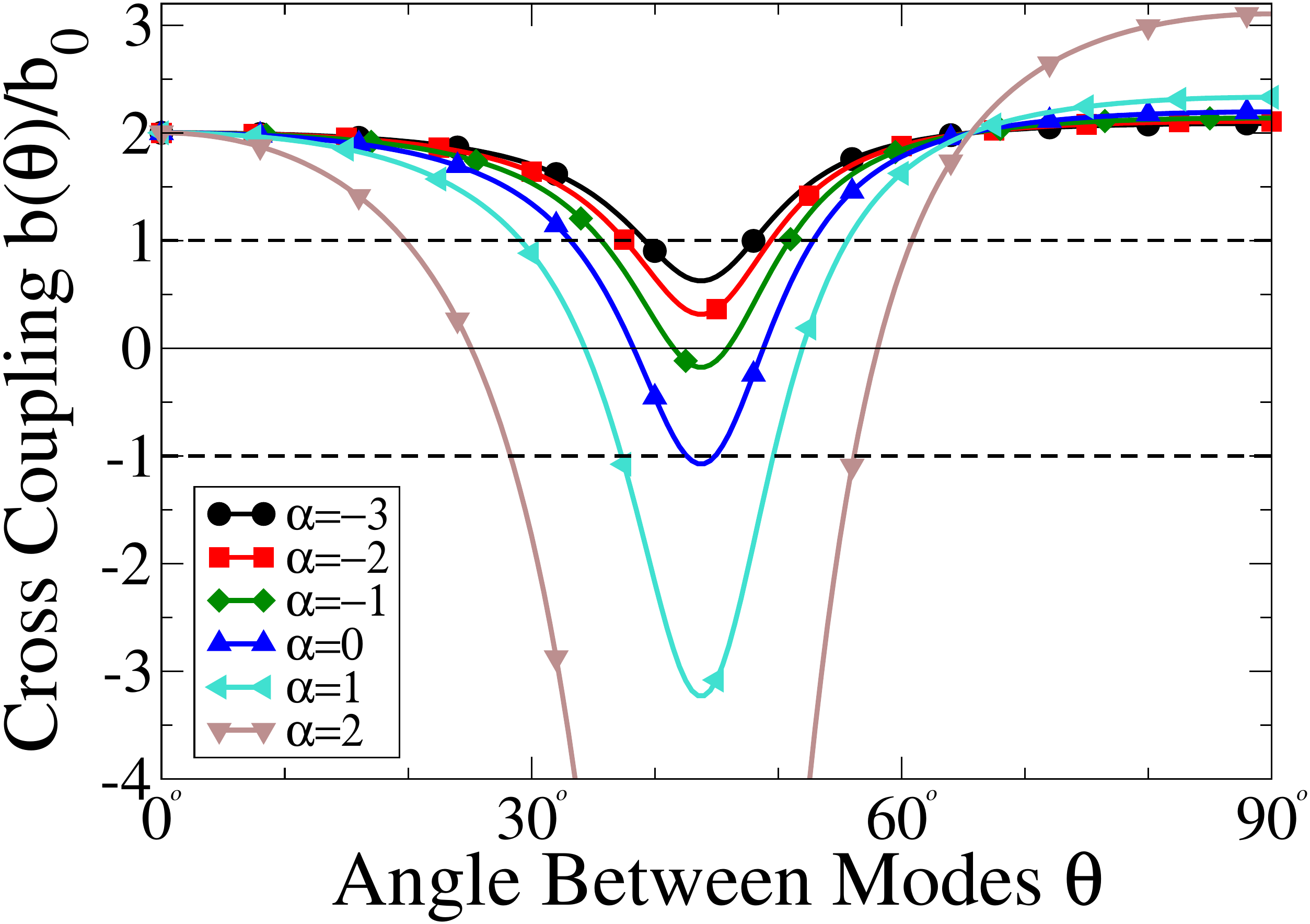}
\par\end{centering}

\caption{Coupling coefficient ratios for $K=2\cos(\tan^{-1}(2/5))$
and $\Phi=\pi/4$ with parameters as in Fig.\ref{neutral1}a for (a)
nonlinear dispersion parameter $\alpha=-1$ and different strengths
of 1:3-forcing $\rho$ and (b) 1:3 forcing strength $\rho=1$ and
different values of the nonlinear dispersion parameter $\alpha$.
\label{fig:Cubic Coefs K is 1.857_phipio4}}
\end{figure}

For $K=2\cos(\tan^{-1}(2/5))$ the competition between modes
subtending an angle near $\theta_{r}$ is strongly enhanced for values
of the  1:3-forcing phase near $\Phi=3\pi/4$ resulting in a suppression
of $\theta_{r}-$rectangles. For the same $K$ but $\Phi=\pi/4,$
we expect correspondingly a selective enhancement of the $\theta_{r}-$rectangles
as long as $\rho$ is sufficiently small so that $b_{0}>0$ (see Fig.\ref{Fig:SelfCoupling}).
This is  shown in Fig.\ref{fig:Cubic Coefs K is 1.857_phipio4}, which
 depicts the behaviour of the cubic coupling coefficient ratio as
a function of the angle $\theta$ for different values of the 1:3-forcing
strength $\rho$ and of the nonlinear dispersion $\alpha$ in this
regime. As the 1:3-forcing strength $\rho$ is increased, and as the
nonlinear dispersion $\alpha$ becomes more positive, the dip with
minimum at $\theta=\theta_{r}=2\tan^{-1}(2/5)$ becomes more
and more pronounced. In parallel the range in $\theta$ of stable
rectangle patterns increases around $\theta_{r}$. Due to the narrowness
of the dip we expect at most stripe and $\theta_{r}$-rectangle patterns
to be stable.  For values of $\rho$ and $\alpha$ such that $b(\theta)/b_{0}<-1$
the corresponding rectangles pattern become subcritical and our weakly
nonlinear analysis to cubic order is insufficient to make predictions
about pattern selection involving those modes.

\subsection{Competition between complex patterns\label{sub:Energy}}

If multiple patterns are simultaneously linearly stable they can coexist
and compete in sufficently large systems. Typically they form then
domains and the competition involves the motion of walls separating
the domains. If the system allows a Lyapunov functional this competition
can be characterized in terms of the difference between the energies
associated with the respective patterns. Unless the domain walls become
pinned by the underlying pattern \cite{Po86,BoVi02a}, it is expected
that the final state arising from random initial conditions consists
of the pattern with minimal energy \cite{ChVi99,Mu93}. 

Because $b(\theta_{ij})=b(\theta_{ji})$, (\ref{eq:GenAmpEqs}) can
indeed be derived from a Lyapunov functional $\mathcal{F}$, such
that $\frac{\partial Z_{j}}{\partial T}=-\frac{\partial\mathcal{F}}{\partial\bar{Z}_{j}^{\ }}$
with \begin{equation}
\mathcal{F}=\sum_{n=1}^{N}\left[-\lambda(\gamma-\gamma_{c})|Z_{n}|^{2}-\frac{1}{2}(b_{0}|Z_{n}|^{4}+\sum_{m=1,m\neq n}^{N}b(\theta_{mn})|Z_{n}|^{2}|Z_{m}|^{2}\right]\label{Lyapunov}\end{equation}
 for a pattern with $N$ modes. The Lyapunov function gives the energy
$\mathcal{F_{N}}$ for an equal-amplitude $N-$mode pattern, which
we rescale to obtain $\hat{\mathcal{F}_{N}}$, \begin{equation}
\hat{\mathcal{F}_{N}}\equiv\frac{\mathcal{F}_{N}}{(\gamma-\gamma_{c})^{2}}=\frac{-\lambda^{2}}{b_{0}+\sum_{n=1,n\neq j}^{N}b(\theta_{jn})},\label{Lyapunov2}\end{equation}
for some $j\in[1,N]$. The sum in the denominator represents the sum
of all cubic coefficients in (\ref{eq:GenAmpEqs}). Since the modes
are evenly spaced, the choice of $j$ in (\ref{Lyapunov2}) is arbitrary.
We use the energy $\mathcal{F_{N}}$ as a guide to predict which $N$-mode
pattern will ultimately emerge. 

\begin{figure}[h!]
\begin{centering}
(a) \includegraphics[width=2.5in]{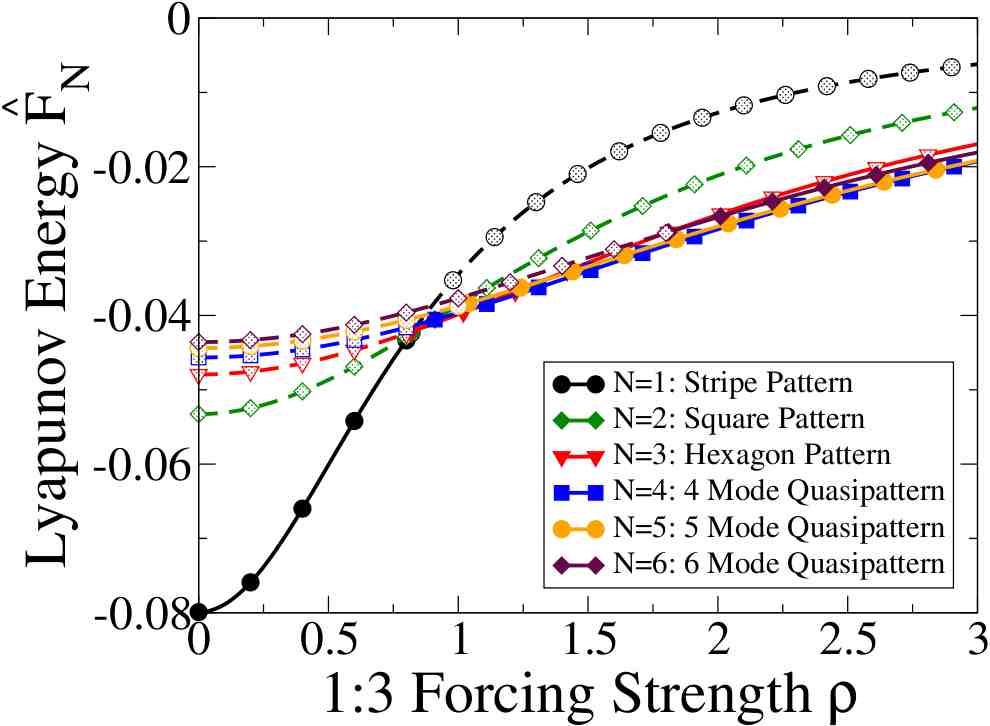}\,\,\,\,\,\,\,\,(b)
\includegraphics[width=2.5in]{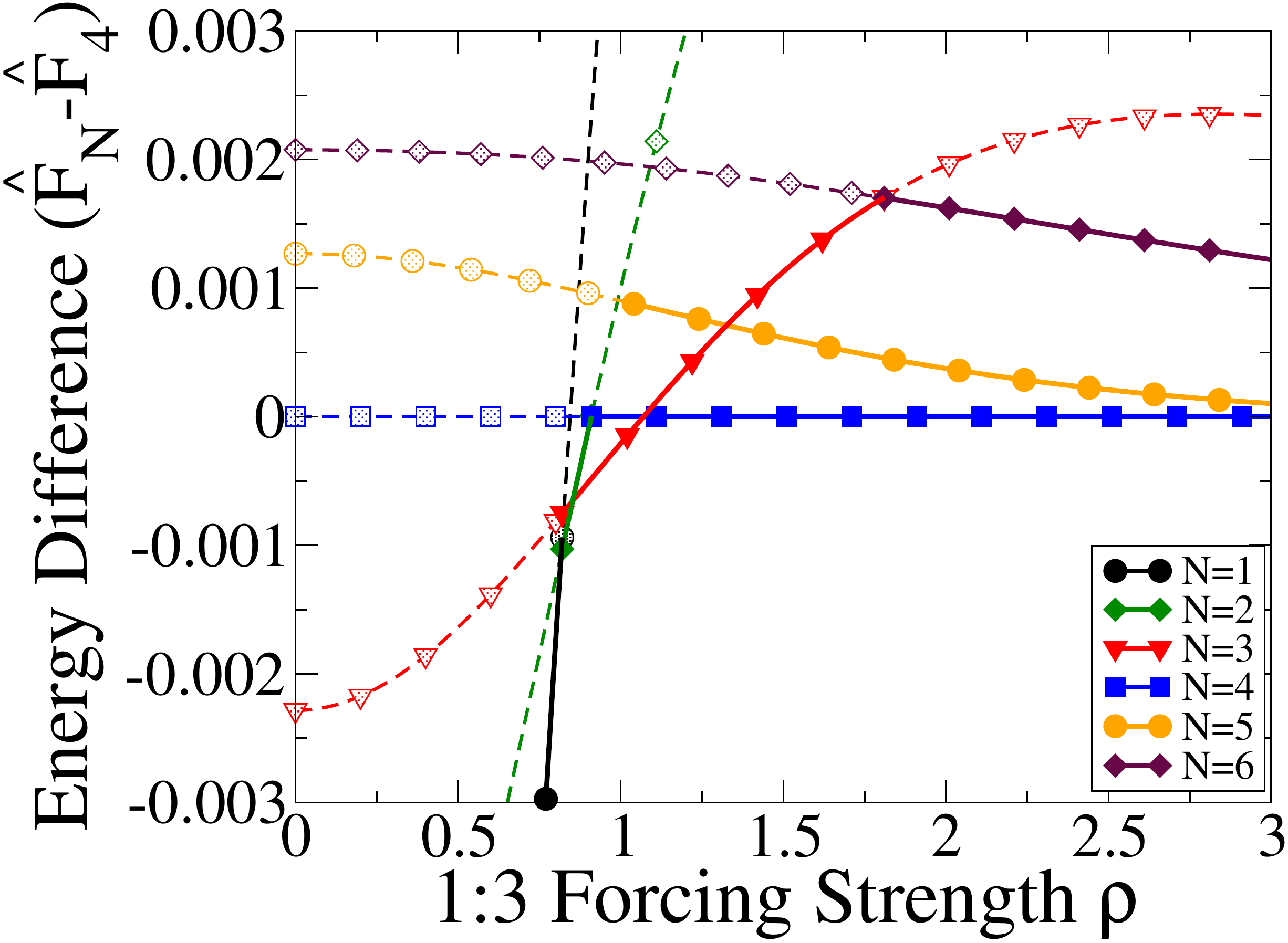}
\par\end{centering}

\caption{(a) Rescaled energies $\hat{\mathcal{F}_{N}}$ for evenly spaced
modes in the case $K=2$. Parameters as in Fig.\ref{neutral1}b with
$\alpha=-1$ and $\Phi=3\pi/4$. Solid (dashed) lines denote linearly
stable (unstable) patterns. (b) Energy difference $\hat{\mathcal{F}_{N}}-\hat{\mathcal{F}_{4}}$
for the same data as shown in (a). The 4-mode pattern is preferred
for $\rho>1.2$. \label{fig:Energy_Kis2}}
\end{figure}

\begin{figure}[h!]
\begin{centering}
\includegraphics[width=2.5in]{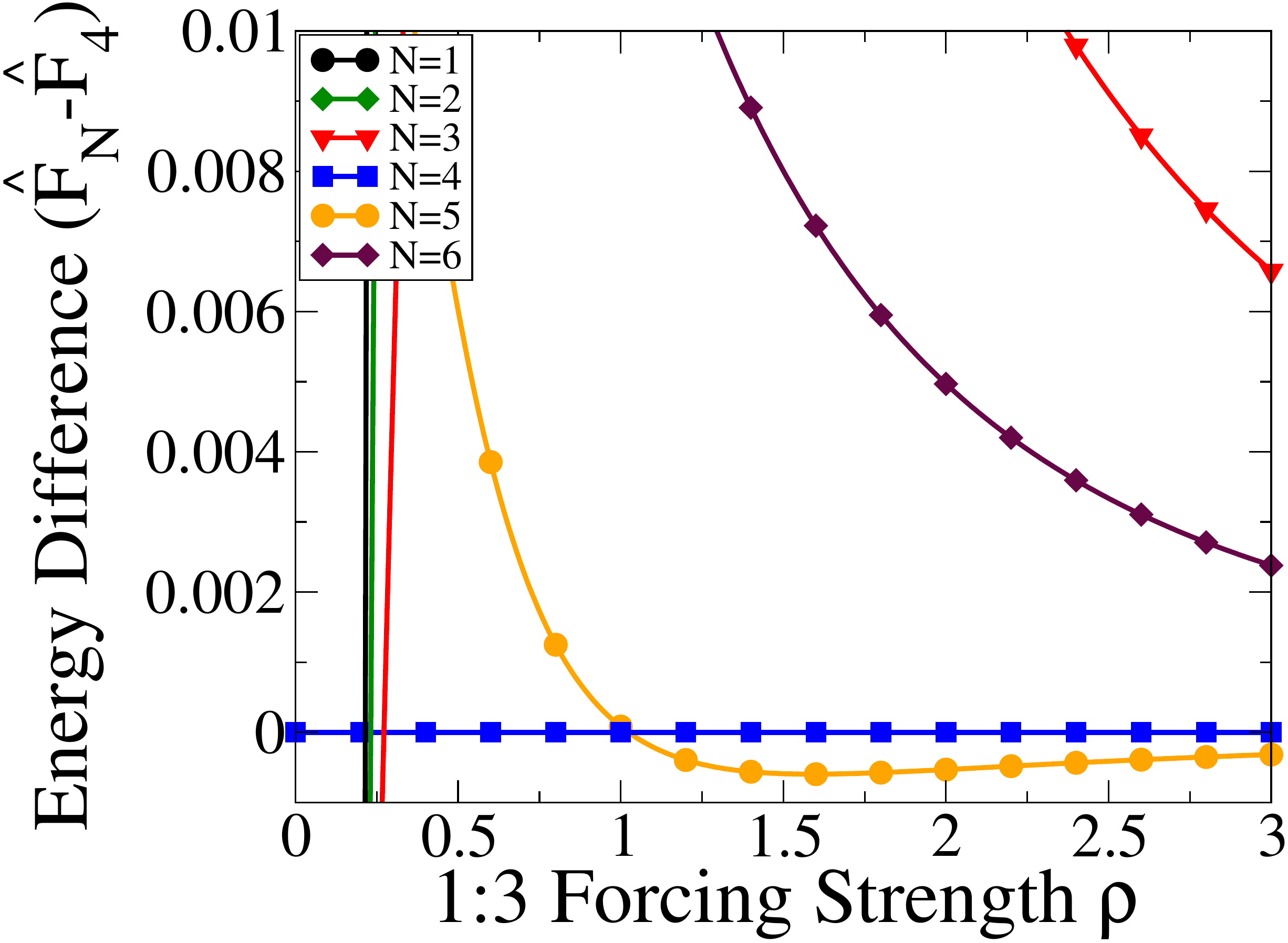}
\par\end{centering}

\caption{Rescaled energies $\hat{\mathcal{F}_{N}}$ for evenly spaced modes
in the case $K=2$. Parameters as in Fig.\ref{neutral1}b with $\alpha=2.5$
and $\Phi=3\pi/4$. 5-mode pattern preferred for $\rho>1$. \label{fig:Energy_Kis2_ais2p5}}
\end{figure}

\begin{figure}[h!]
\begin{centering}
(a) \includegraphics[width=2.5in]{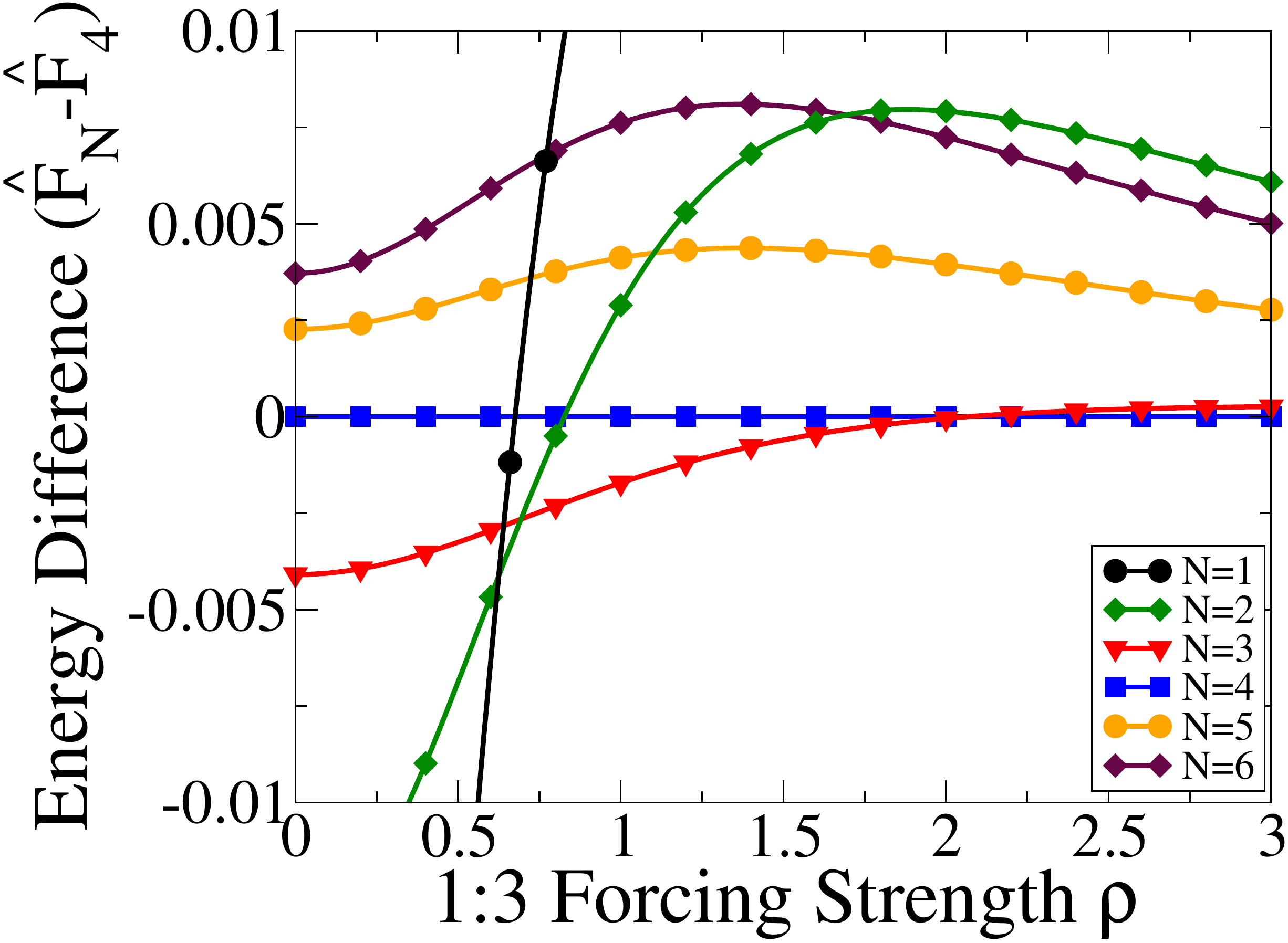}\,\,\,\,\,\,\,\,(b)
\includegraphics[width=2.5in]{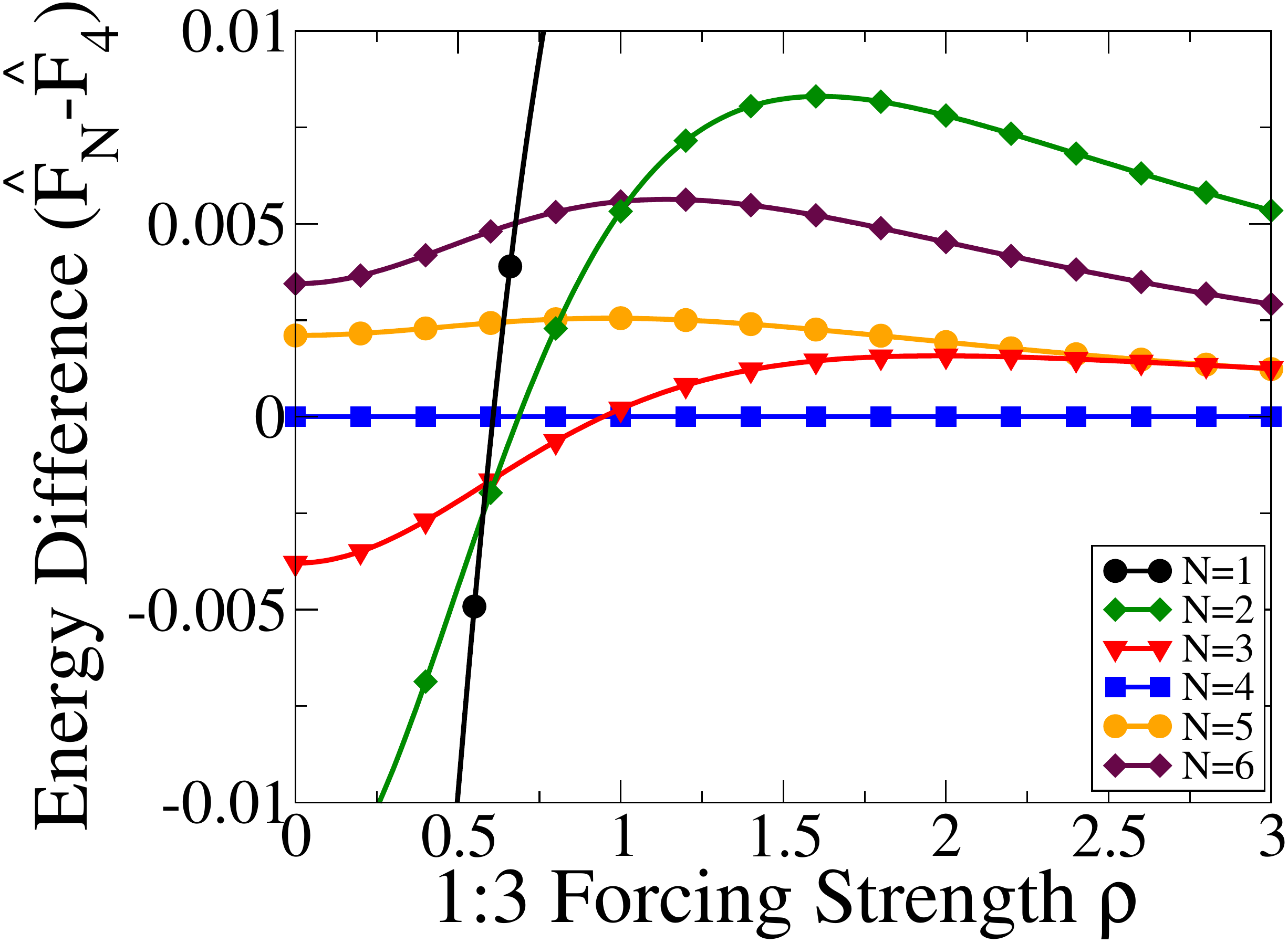}
\par\end{centering}

\caption{\label{fig:Energies_sigmaIs4and6}Pattern selection for Belousov-Zhabotinsky
parameters \cite{HySo93}. Rescaled energy difference $\hat{\mathcal{F}_{N}}-\hat{\mathcal{F}_{4}}$
for $N$ evenly spaced modes. Here $\alpha=-0.4,$ $\beta=1.4$ ,
$\sigma=4$ (a) and $\sigma=6$ (b). }
\end{figure}

\begin{figure}[h!]
\begin{centering}
(a) \includegraphics[width=2.5in]{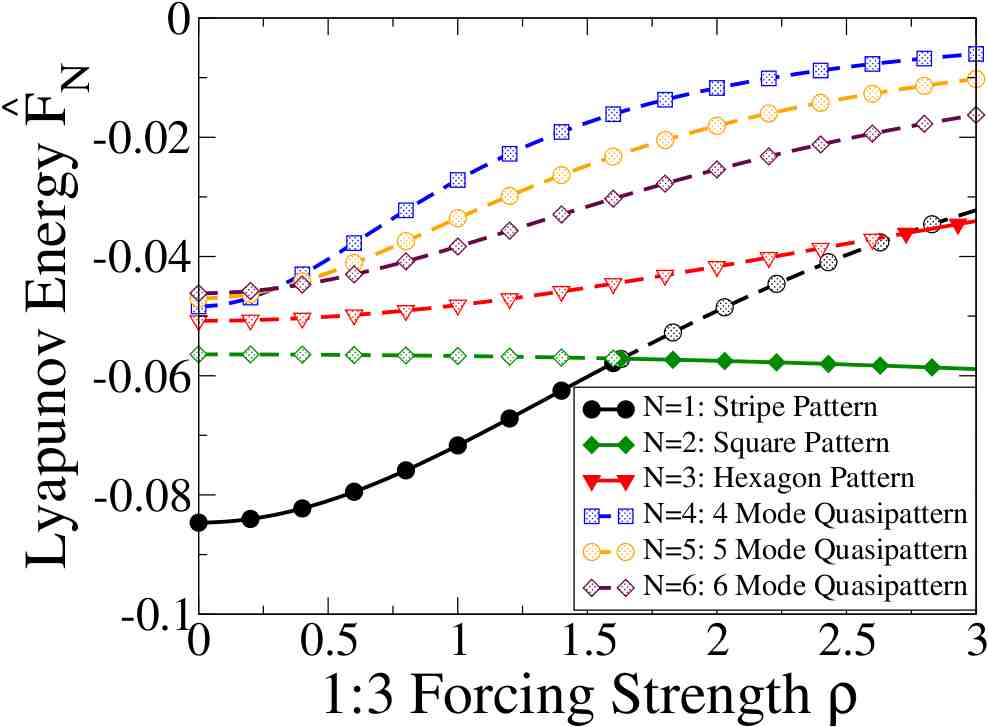}\,\,\,\,\,\,\,\,(b)
\includegraphics[width=2.5in]{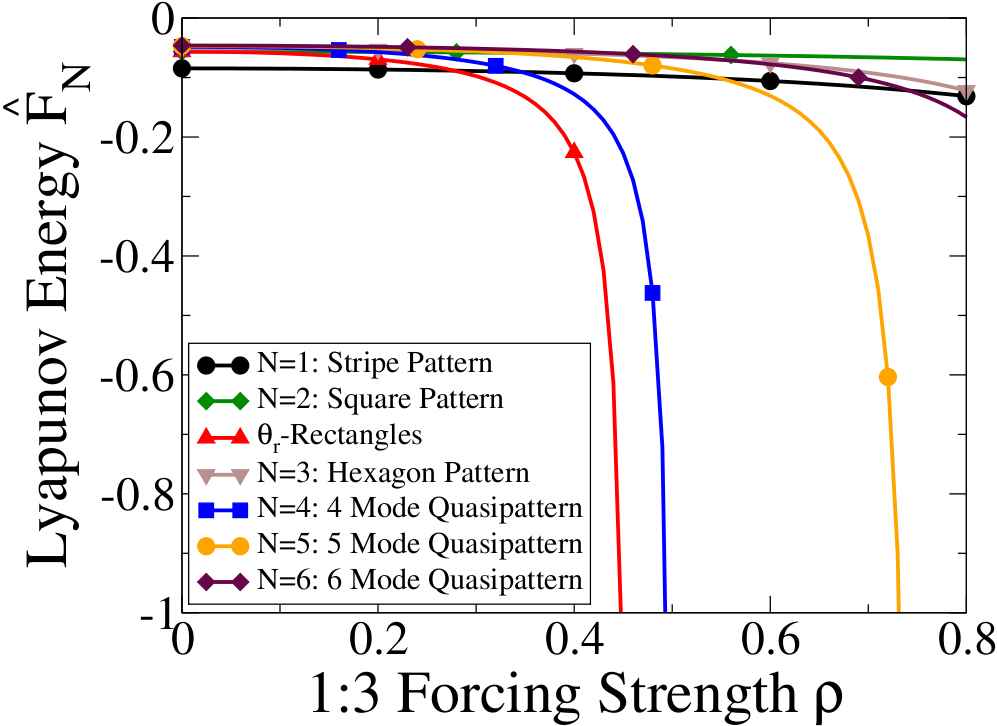}
\par\end{centering}

\caption{Rescaled energies $\hat{\mathcal{F}_{N}}$ for evenly spaced modes
in the case $K=2\cos(\tan^{-1}(2/5))$. Parameters as in Fig.\ref{neutral1}a
with $\alpha=-1$ and (a) $\Phi=3\pi/4$ (cf. Fig.\ref{fig:Cubic Coefs K is 1.857_phi3pio4}),
(b) $\Phi=\pi/4$ (cf. Fig.\ref{fig:Cubic Coefs K is 1.857_phipio4}).
In (a), solid (dashed) lines denote linearly stable (unstable) patterns.
\label{Fig:Energies - K is 1p857} }
\end{figure}

Figs.\ref{fig:Energy_Kis2}-\ref{Fig:Energies - K is 1p857} show
how the energies $\hat{\mathcal{F_{N}}}$ vary with $\rho$, the strength
of forcing near 1:3 resonance, for different patterns with $N$ modes
evenly spaced in Fourier space. Since the energy depends smoothly
on the angles $\theta_{jn}$ (cf. (\ref{Lyapunov2})) little change
in the energy is expected if the modes are not quite evenly spaced. 

For $K=2$ and $\alpha=-1$ the energies of the patterns with more
than 3 modes are very close to each other, as seen in Fig.\ref{fig:Energy_Kis2}a.
For clarity we show therefore in Fig.\ref{fig:Energy_Kis2}b the difference
$\mathcal{\hat{\mathcal{F}}}_{N}-\hat{\mathcal{F}}_{4}$. As one increases
$\rho$, first stripe patterns ($N=1$), then square patterns ($N=2$),
then hexagons ($N=3$), and eventually supersquares ($N=4$) have
the lowest energy. For the respective parameter values, these are
therefore the patterns we expect eventually to arise from noisy initial
conditions. Also indicated in Figs.\ref{fig:Energy_Kis2}a,b is the
linear stability of these patterns. Here we determine the linear stability
of an $N$-mode pattern by calculating the linear stability of that
pattern within the space spanned by the $N$ evenly spaced modes and
an additional mode rotated an arbitrary angle $\psi$ with respect
to the ${\bf \textbf{k}}_{1}$-mode in Fourier space. For larger $\rho$
patterns comprised of $5$ and $6$ modes are linearly stable but
do not have the lowest energy, though for $\rho>2$ the energies of
the patterns with 4 and 5 modes are very close.

In Sec.\ref{sub:Amplitude-Equations} we showed that for a more positive
$\alpha$ relatively small values of the 1:3-forcing strength $\rho$
 are sufficient to stabilize rectangle patterns over a wide range
of $\theta$  (cf. Fig. \ref{fig:alpha-dependence}a). As a result
we expect that the resonant triad can stabilize patterns comprised
of more modes than was possible in the case $\alpha=-1$ depicted
above. Fig.\ref{fig:Energy_Kis2_ais2p5}b shows that this is indeed
the case. Increasing $\alpha$ to $\alpha=2.5$ reduces for $\rho>1$
the energy of the 5-mode patterns below that of the 4-mode pattern
and renders it the pattern with the lowest energy among the patterns
with equally spaced modes. We therefore expect that in numerical simulations
5-mode patterns would arise over a large range in $\rho$.

For the parameter set that is relevant for the Belousov-Zhabotinsky
reaction \cite{HySo93} we found that the mode competition is also
sufficiently reduced to suggest the stability of multi-mode patterns
(cf. Fig.\ref{fig:ThetaVsRho_CrossCouplingIs1}). This is confirmed
in Fig.\ref{fig:Energies_sigmaIs4and6}, which shows the energies
for $\alpha=-0.4$ and $\beta=1.4$. While for $\sigma=4$ the 4-mode
pattern has an energy that is only barely below that of the hexagons
and reaches those lower values only for large forcing strengths $\rho$
(Fig.\ref{fig:Energies_sigmaIs4and6}), increasing $\sigma$ to $\sigma=6$
pushes the energy of the 4-mode pattern well below that of the hexagons
(Fig.\ref{fig:Energies_sigmaIs4and6}b). We therefore expect that
the complex patterns we discuss in this paper are accessible in this
experimental system.

For $K=2\cos(\tan^{-1}(2/5))$ with $\Phi=3\pi/4$
(Fig.\ref{Fig:Energies - K is 1p857}a), stripes are again stable
for small $\rho.$ As $\rho$ is increased stripes lose stability
and square patterns ($N=2$) become linearly stable and have the lowest
energy beyond $\rho\approx1.6$. Hexagon patterns ($N=3$) become
linearly stable at $\rho=2.7$, but do not have lower energy than
the square patterns. Patterns with more than 3 modes are linearly
unstable, which is consistent with the predictions based on the coupling
coefficient shown in Fig.\ref{fig:Cubic Coefs K is 1.857_phi3pio4}a:
Fourier modes subtending angles smaller than $\pi/3$ compete strongly
and therefore do not coexist stably. In simulations starting from
random initial conditions, we therefore expect either stripe patterns
(for $\rho<1.6$) or square patterns to arise. For $\Phi=\pi/4$,
shown in Fig.\ref{Fig:Energies - K is 1p857}b, stripes again are
stable and have the lowest energy for small $\rho$. For $\rho=0.28$
they become unstable to rectangle patterns spanned by modes subtending
an angle of $\theta=\theta_{r}\equiv2\tan^{-1}(2/5)$. As
$\rho$ is increased to $\rho=0.46$ this rectangle pattern becomes
subcritical (cf. Fig.\ref{fig:Cubic Coefs K is 1.857_phipio4}) and
the weakly nonlinear analysis taken to cubic order and the associated
energy arguments are not sufficient to make predictions about pattern
selection. For $\rho>0.28$  none of the patterns with equally spaced
modes  has lower energy than the rectangle pattern selected by the
resonant triad. Note that in the cubic truncation (\ref{Lyapunov2})
the energy for a pattern diverges when the respective bifurcation
changes direction. 

\section{Numerical Simulations in Large Domains\label{sec:Results}}

To  confirm our predictions for the pattern selection, we perform
numerical simulations of the complex Ginzburg-Landau equation (\ref{eq:cgle_final}).
Being interested in the formation of complex patterns comprised of
3 or more modes, we focus on the case $K=2$. The linear parameters
are as in Fig.\ref{neutral1}a with nonlinear parameters $\alpha=-1$,
$\Phi=3\pi/4$, and various values of $\rho$. We use periodic
boundary conditions and employ a pseudo-spectral method with Crank-Nicolson-Adams-Bashforth
time stepping 

\begin{figure}[h!]
\begin{centering}
\includegraphics[width=6in]{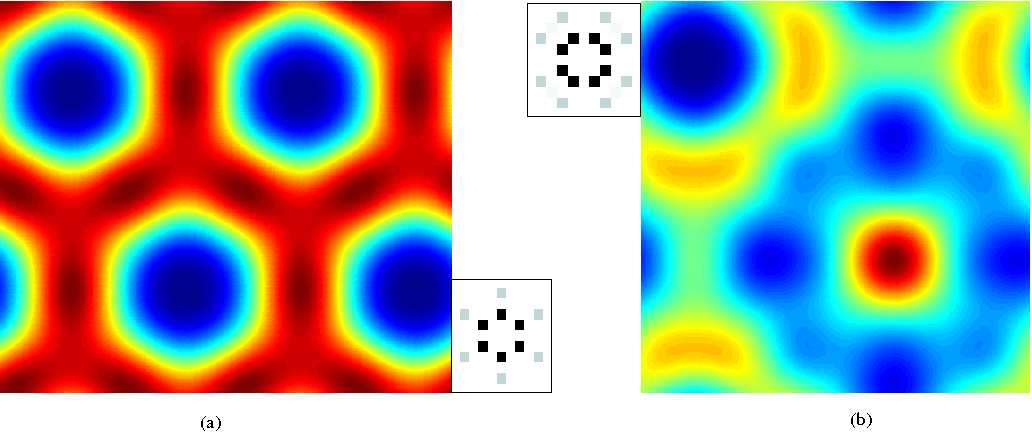}
\par\end{centering}

\caption{Small-system numerical simulations of (\ref{eq:cgle_final})
  for 1:3 forcing strength $\rho=1$. (a) Hexagon pattern in
  rectangular domain, $L_{x}=4\pi/k^{SH}\approx 23.49$,
  $L_{y}=L_x/\sqrt{3}\approx 13.56$. (b) Supersquare pattern in square
  domain, $L=2\pi/(\cos(\tan^{-1}(1/2))k^{(SH)}\approx
  13.13$. Parameters as in Fig.\ref{neutral1}b (case $K=2$) with
  $\alpha=-1$, $\Phi=3\pi/4$. \label{fig:SmallboxSims}}
\end{figure}

First, to test the weakly nonlinear analysis we focus on the regime
where both hexagons and 4-mode patterns are linearly stable (Fig.\ref{fig:Energy_Kis2})
and run numerical simulations in domains of minimal size for each
pattern. Thus, all participating modes lie exactly on the critical
circle. Since the 4-mode pattern with evenly spaced modes does not
lie on a regular Fourier grid we approximate it by modes spaced at
$\theta_{1}=2\tan^{-1}(1/2)\approx53^{\circ}$ and $\theta_{2}=\pi/2-2\tan^{-1}(1/2)\approx37^{\circ}$
apart. Due to the smooth dependence of the energy on the angles $\theta_{jn}$
(cf. Eq.(\ref{Lyapunov2})) little change in the energy is expected
if the modes are not quite evenly spaced. From Fig.\ref{fig:Energy_Kis2}b,
for a 1:3-forcing strength $\rho=1$ hexagons and 4-mode patterns
are both linearly stable to stripes and the 4-mode patterns are also
linearly stable to square patterns. The numerical simulations confirm
these stability predictions. Starting with a noisy initial condition
generated from a uniform distribution with amplitude 0.05 we get as
expected hexagons (Fig.\ref{fig:SmallboxSims}a) and supersquares
(Fig.\ref{fig:SmallboxSims}b). Fig.\ref{fig:SmallboxSims} also illustrates
the slight diffference in the domain sizes used in the simulations,
which are required to accomodate the patterns. 

To investigate the competition between $N$-mode patterns with
different values of $N$ in the same computational domain we perform
simulations in a large system of linear size $L_{x}=L_{y}\equiv L$,
given below, representing 40 wavelengths. Fig.\ref{fig:Energy_Kis2}b
shows that near $\rho=1$ hexagons and 4-mode patterns are both
linearly stable, but the pattern with minimal energy depends on
$\rho$. In order to investigate the competition between these
planforms we start each simulation with the same noisy initial
condition, generated from a uniform distribution with amplitude 0.01,
and vary only $\rho$. Moreover, in order to clearly identify the
\emph{nonlinear} competition between hexagons and 4-mode patterns and
its dependence on $\rho$, we choose $L$ such that the Fourier modes
with $\theta=\pi/3$ and $\theta=\pi/4$ have the same growth
rates. This is achieved by taking
$L=40(2\pi/(k^{(SH)}-d))\approx473.39$ for which the modes for the
hexagons and the 4-mode patterns are at an equal distance $d=0.004159$
on opposite sides of the critical circle.

\begin{figure}[h!]
\begin{centering}
(a) \includegraphics[width=2.2in]{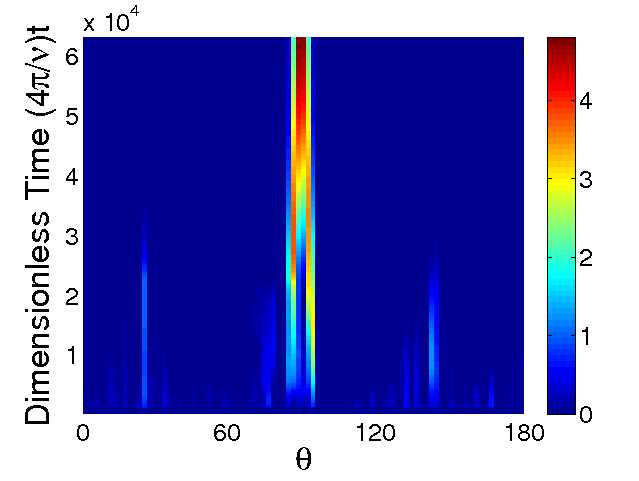}\,\,\,\,\,\,\,\,(b)
\includegraphics[width=2.2in]{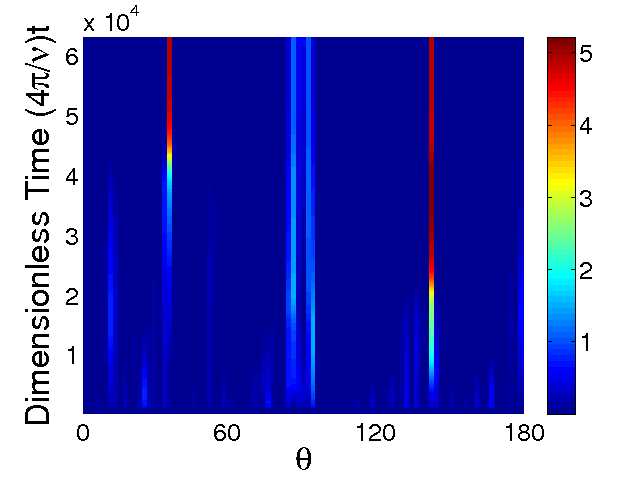}
\par\end{centering}

\begin{centering}
(c) \includegraphics[width=2.2in]{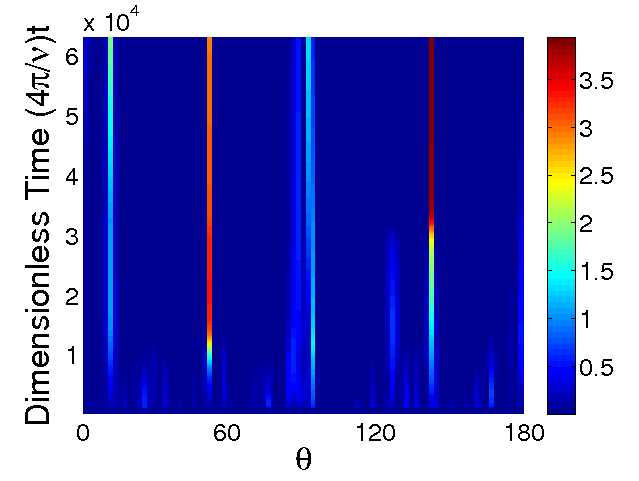}\,\,\,\,\,\,\,\,(d)
\includegraphics[width=2.2in]{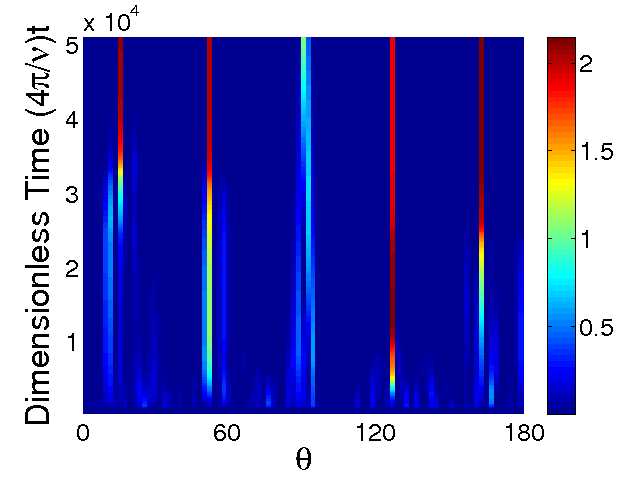}
\par\end{centering}

\begin{centering}
(e) \includegraphics[width=2.2in]{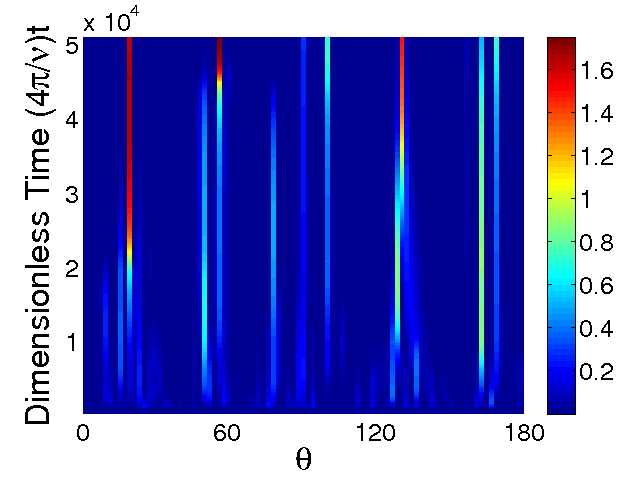}
\par\end{centering}

\caption{Strobed  time dependence of the power spectrum in an annulus around
the critical circle for varying 1:3 forcing strengths $\rho$. Parameters
as in Fig.\ref{neutral1}b with $\alpha=-1$ and $\Phi=3\pi/4$. All
simulations use the same noisy initial condition. (a) $\rho=0.8$,
(b) $\rho=1$, (c) $\rho=1.2$, (d) $\rho=2$, (e) $\rho=3$.\label{fig:TimePlots}}
\end{figure}

\begin{figure}[h!]
\begin{centering}
(a) \includegraphics[width=1.4in]{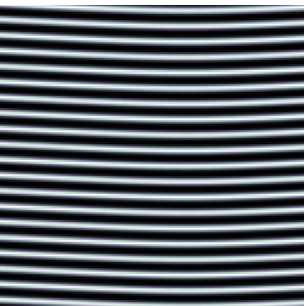}\,\,\,\,\,\,\,\,(b)
\includegraphics[width=1.4in]{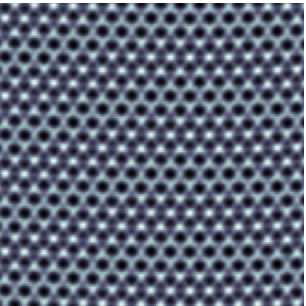}\,\,\,\,\,\,\,\,(c)
\includegraphics[width=1.4in]{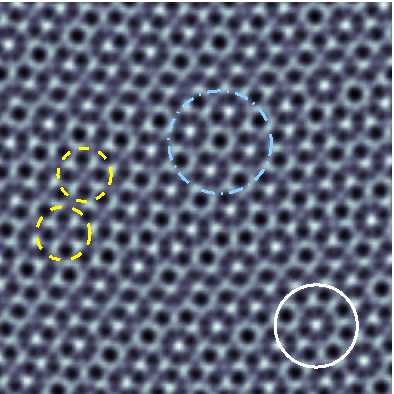}
\par\end{centering}

\begin{centering}
(d) \includegraphics[width=1.4in]{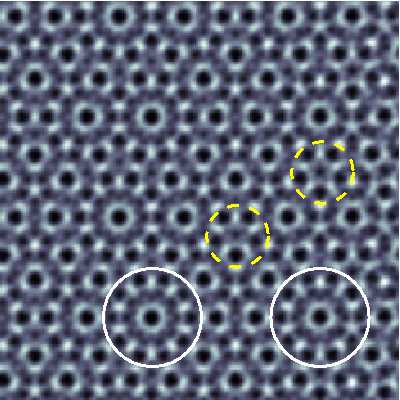}\,\,\,\,\,\,\,\,(e)
\includegraphics[width=1.4in]{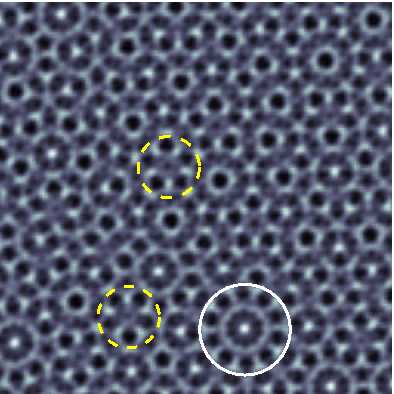}
\par\end{centering}

\caption{Zoom view ($0.25L\times0.25L$) of final state at time $4\pi t/\nu=5\cdot10^{4}$
for varying values of 1:3 forcing $\rho$ with $L=40(2\pi/(k^{(SH)}-d))\approx473.39$.
Other parameters as in Fig.\ref{neutral1}b (case $K=2$) with $\alpha=-1,$
$\Phi=3\pi/4$. The circles mark characteristic features of the patterns.
(a) $\rho=0.8$, (b) $\rho=1$, (c) $\rho=1.2$, (d) $\rho=2$, (e)
$\rho=3$. \label{fig:Solutions}}
\end{figure}

To visualize the competition between the modes near the critical circle,
Fig.\ref{fig:TimePlots} shows the temporal evolution of the magnitude
of the Fourier modes inside a narrow annulus around the critical circle,
divided into bins with an angular width of $2^{\circ}$ plotted stroboscopically
at multiples of the period $4\pi/\nu$. For $\rho=0.8$, based on
energy arguments and the linear stability illustrated in Fig.\ref{fig:Energy_Kis2}b,
we anticipate the final state to consist of a stripe pattern. This
is confirmed by the numerical simulation. While in the Fourier transform
Fig.\ref{fig:TimePlots}a initially three modes come up, reflecting
the linear stability of hexagons for $\rho=0.8$, ultimately the pattern
with lower energy, stripes, dominates and only a single mode remains.
The competition can also be seen in in TimeEvolutionMovie\_rhoIs0p8.mov,
which shows the temporal evolution of the full pattern strobed with
period $4\pi/\nu$. More careful inspection shows that the middle
peak, which seems weaker, actually consists of two modes of lesser
strength that are slowly converging to a single strong peak. The splitting
of the peak into two modes reflects a slight undulation of the resulting
pattern, which is apparent in the snapshot of part of the final solution
shown in Fig.\ref{fig:Solutions}a.

Fig.\ref{fig:TimePlots}b shows the evolution of the Fourier transform
for $\rho=1$ corresponding to the time evolution of the full pattern
shown in TimeEvolutionMovie\_rhoIs1.mov. As with the $\rho=0.8$ case,
initially the amplitudes of three modes grow. In contrast to the $\rho=0.8$
case, however, for $\rho=1$ stripes are unstable to hexagons, and
hexagons have the lowest energy. Correspondingly the three modes persist
resulting in the hexagon patterns shown in Fig.\ref{fig:Solutions}b.
Note that the hexagon patterns are comprised of domains of up- and
down-hexagons characterized by white and black centers, respectively.
This reflects the fact that the amplitude equations (\ref{eq:GenAmpEqs})
for these subharmonic patterns have no quadratic term, so neither
the up- nor the down-hexagons are preferred. Whether eventually one
of the two types wins out over the other depends on the interaction
between the fronts connecting the domains.

For $\rho=1.2$ we anticipate from Fig.\ref{fig:Energy_Kis2}b that
the resulting solution will be a 4-mode pattern. Indeed, Fig.\ref{fig:TimePlots}c
shows 4 modes in the Fourier transform, though they are not quite
equally spaced. The corresponding pattern evolution is shown in TimeEvolutionMovie\_rhoIs1p2.mov;
Fig.\ref{fig:Solutions}c shows the final state,  characterized by
supersquare (dash-dotted, blue circle) and antisquare (dashed, yellow circles)
elements with approximate 4-fold rotational symmetry \cite{DiSi97},
as well as  approximate 8-fold symmetric elements (solid, white
circle). 

From Fig.\ref{fig:Energy_Kis2}b a 4-mode pattern is also anticipated
for $\rho=2$, but the numerical simulation actually results in a
5-mode pattern as shown in Fig.\ref{fig:TimePlots}d. The energies for
evenly-spaced 4- and 5-mode patterns are very close for this forcing
strength. Therefore changes in the energies that result from an uneven
distribution of the modes may render the 5-mode pattern energetically
lower. The corresponding pattern features elements with approximately
5- and 10-fold rotational symmetry (dashed,yellow and solid, white
circles, respectively, in Fig.\ref{fig:Solutions}d). For $\rho=3$ even
more modes persist for a long time and at the final time shown, $4\pi
t/\nu=5\cdot10^{4}$, the pattern has not settled yet into its final
state. The temporal evolution of the full patterns for $\rho=2$ and
$\rho=3$ are shown in TimeEvolutionMovie\_rhoIs2.mov and
TimeEvolutionMovie\_rhoIs3.mov, respectively.

\begin{figure}[h!]
\begin{centering}
\includegraphics[width=3in]{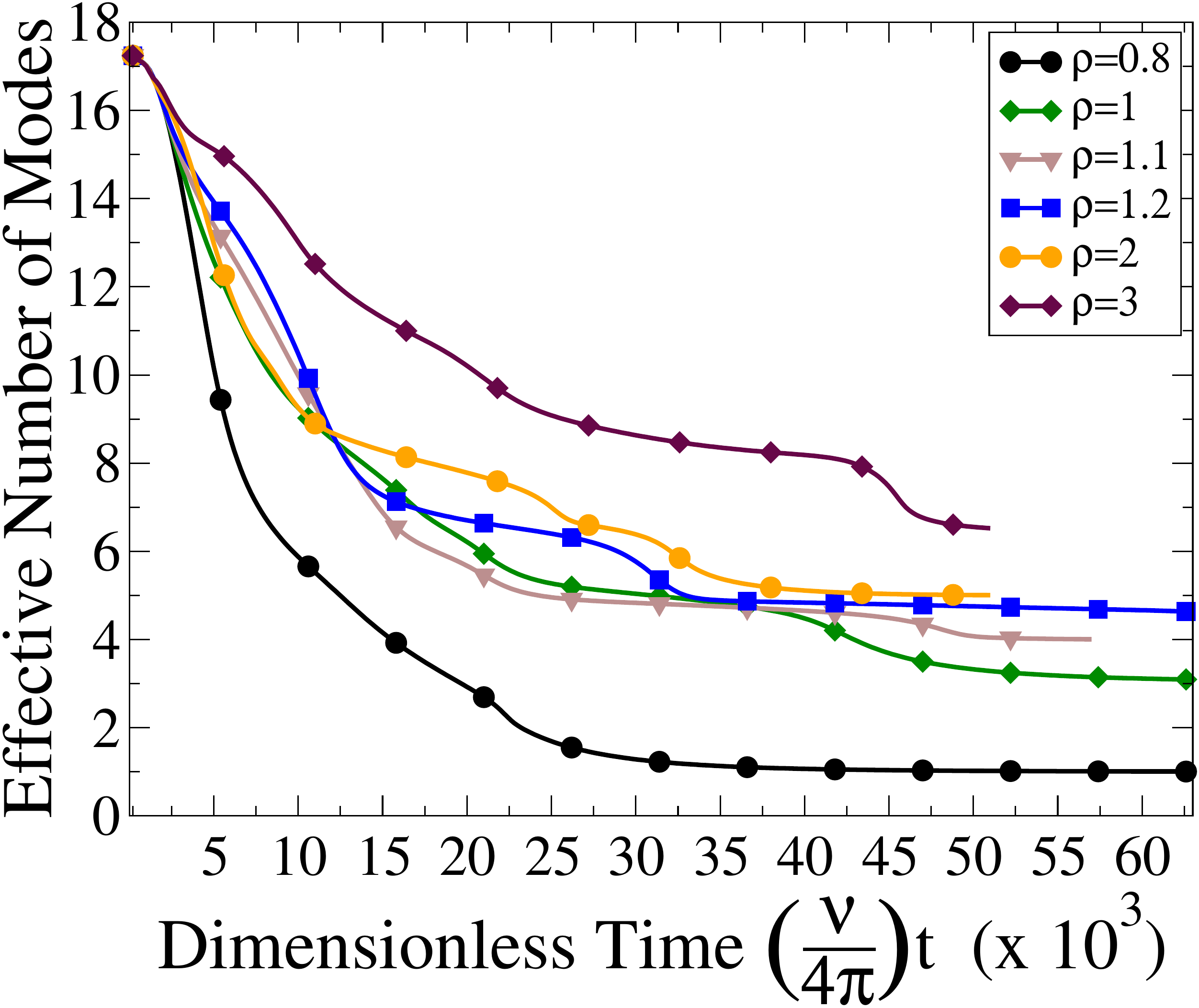} 
\par\end{centering}

\caption{Effective number of Fourier modes $e^{S}$ plotted as a function
of time for different values of the 1:3 forcing strength $\rho$.
Data taken from numerical simulations of (\ref{eq:cgle_final}) for
parameters as in Fig.\ref{neutral1}b (case $K=2$) with $\alpha=-1,$
$\Phi=3\pi/4$. \label{fig:Entropy}}
\end{figure}

To characterize the ordering process taking place in the evolution
of the patterns we calculate the spectral pattern entropy $S=\sum_{i,j}p_{ij}\ln p_{ij}$,
where $p_{ij}=\tilde{p_{ij}}/(\sum_{i,j}\tilde{p}_{ij})$ is
the normalized power in the Fourier mode with amplitude $\tilde{p}_{ij}$
and the sum includes all modes within the critical annulus binned
into $10^{\circ}$ segments. The entropy allows an estimate of the
effective number of Fourier modes $e^{S}$contributing to the pattern.
Fig.\ref{fig:Entropy} illustrates the temporal evolution of $e^{S}$
for different values of the 1:3 forcing strength $\rho$. The number
of significant modes increases with $\rho,$ although not monotonically
for all times: for example, the curves corresponding to $\rho=1$
and $\rho=1.2$ cross several times, near dimensionless times $\nu t/4\pi=12\times10^{3},17\times10^{3},35\times10^{3}$,
before beginning to converge smoothly to $e^{S}\approx3$ and $4$,
respectively. While the decrease in the effective number of Fourier
modes with time in the transients is monotonic it occurs in spurts.
For $\rho=3$, for instance, there is a sudden dip near dimensionless
time $\nu t/4\pi=45\times10^{3}$. These dips are related to the sudden
disappearance of modes that is apparent in the time evolution plots
Fig.\ref{fig:TimePlots}a-e.

\begin{figure}[h!]
\begin{centering}
\includegraphics[height=2in]{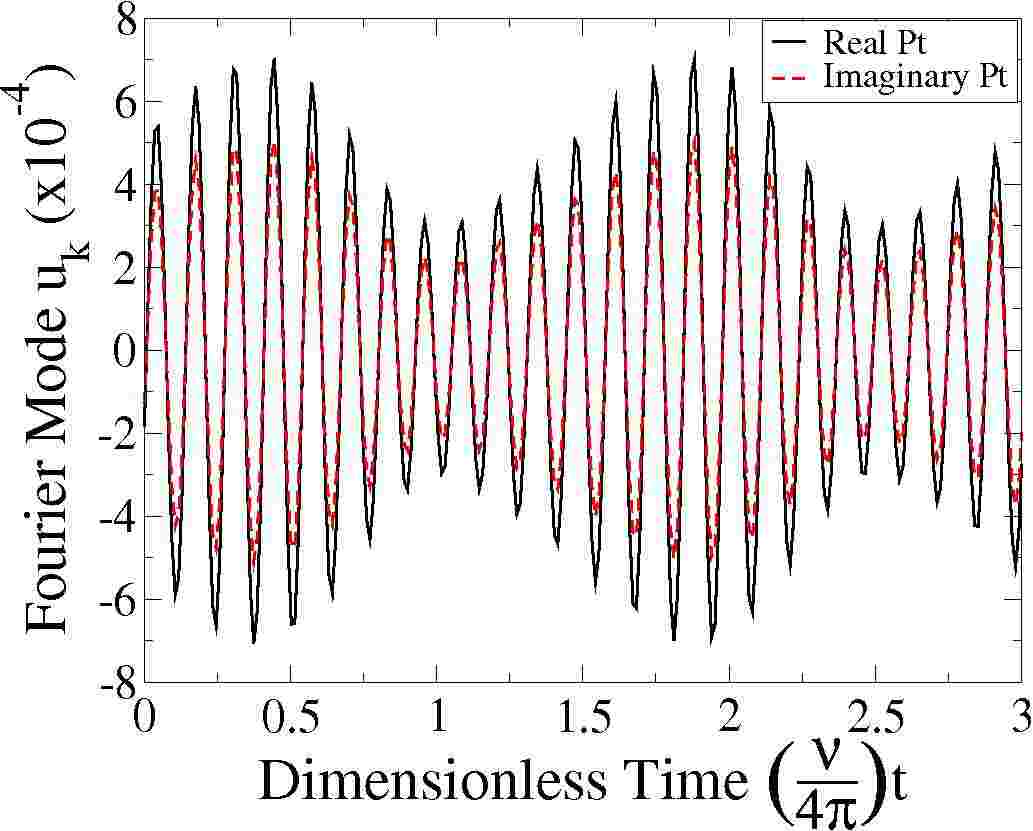}
\par\end{centering}

\caption{ Temporal evolution of the real and imaginary parts of one of the
spatial Fourier modes of the pattern shown  in Fig.\ref{fig:Solutions}e
and in the movie SimulationOver2Periods\_withFastOscs.mov.  \label{fig: u movie}}
\end{figure}

The complex Ginzburg-Landau equation (\ref{eq:cgle_final}) describes
the evolution of the oscillation amplitude on a slow time. The full
time evolution of the underlying system involves also the Hopf frequency.
To illustrate the full time dependence the movie SimulationOver2Periods\_withFastOscs.mov shows the evolution of the pattern over two forcing periods obtained
for $\rho=2$ as one might see it in an experiment. Fig.\ref{fig: u movie}
shows the corresponding temporal evolution of the real and imaginary
parts of one of the spatial Fourier modes. The beating reflects the
quasi-periodic forcing of the system. Details about the quantity shown
 in the movie SimulationOver2Periods\_withFastOscs.mov and in Fig.\ref{fig: u movie}
are given in the Appendix.

\section{Conclusion}

\label{sec:Conclusion}

We have demonstrated analytically and confirmed numerically that in
systems that are near a Hopf bifurcation to spatially uniform oscillations
forcing with judiciously chosen waveforms can stabilize complex periodic
and quasi-periodic patterns comprised of up to 4 and 5 Fourier modes.
Essential for the success of this approach was the use of a quasi-periodic
forcing function in which the frequency content near twice the Hopf
frequency consists of two frequencies; thus, the forcing is slowly
modulated in time. Since the investigated complex patterns are subharmonic
in time with respect to that modulation they arise in a pitch-fork
bifurcation and are amenable to a weakly nonlinear analysis. The stabilization
of the complex patterns was achieved by exploiting resonant triad
interactions that result from the additional forcing component with
a frequency close to three times the Hopf frequency. 

As expected from investigations in the context of the Faraday system
\cite{ZhVi97,ZhVi97a,ChVi97,ChVi99,SiTo00,PoTo03}, our weakly nonlinear
analysis showed that the resonant triad interaction can significantly
modify the competition between Fourier modes of different orientation
through the excitation of weakly damped modes. This damping is controlled
by the differences in frequency and by the amplitude of the two forcing
components near twice the Hopf frequency. It can be chosen to enhance
the self-coupling of the modes significantly, which effectively reduces
the competition between modes over a quite wide range of angles subtended
by those modes. As in the Faraday system, this stabilizes complex
patterns in the forced oscillatory systems investigated here. Alternatively,
the forcing can be chosen to reduce the competition between modes
differing in their orientation be a quite specific angle $\theta_{r}$,
which can be tuned over quite some range. In the Faraday system this
approach can also lead to complex 4-mode patterns, which effectively
consist of the combination of two square patterns that are rotated
by an angle close to $\theta_{r}$. In the forced oscillatory systems
investigated here this mechanism alone does not yield complex patterns
since in the unforced system squares are never stable, while in the
Faraday system square patterns are stable for low viscosity \cite{KuGo96,ChVi97,ZhVi96}.
Thus, in the present context this mechanism only leads to the stabilization
of rectangle patterns. For sufficiently strong forcing their bifurcation
can even be made subcritical, even though the stripe patterns (and
the Hopf bifurcation itself) are supercritical. 

We have complemented the weakly nonlinear analysis by direct simulations
of the extended complex Ginzburg-Landau equation in large domains
to study the competition between different, linearly stable, complex
patterns. As expected from the variational character of the amplitude
equations and the dependence of the energy of the various patterns
on the forcing strength, we find that as the resonant triad interaction
is increased more complex patterns dominate over simpler patterns.
For the parameters chosen in the numerical simulations we find patterns
with 4-fold symmetric elements reminiscent of super-squares and anti-squares
\cite{DiSi97} as well as 5-fold symmetric elements. The weakly nonlinear
analysis (see Fig.\ref{Fig:Cubic Coefs Kis2}b) suggests that for
other system parameters patterns comprised of more modes yet could
be stable. 

It should be noted that we reached the regime in which complex patterns
are stable by tuning the amplitudes and phases of the forcing function,
which constitute external forcing parameters. Therefore we expect
that the complex patterns should be accessible quite generally in
forced oscillatory systems, in particular also in chemical oscillators
\cite{PeOu97,LiHa04,YoHa02}. Specifically, our weakly nonlinear analysis
indicates that the dependence of the patterns on the nonlinear dispersion
coefficient $\alpha$ of the unforced system can be compensated by
the strength of the forcing close to three times the Hopf frequency
(cf. Fig.\ref{Fig:Cubic Coefs Kis2}b).  Moreover, for the parameters
determined experimentally in the oscillatory Belousov-Zhabotinsky
reaction \cite{HySo93} we have shown explicitly that moderate forcing
strengths are sufficient to stabilize 4-mode patterns. It should be
noted,  however, that the complex patterns possibly arise only very
close to onset. Therefore the application of our results to experiments
may require systems with relatively large aspect ratios and a very
careful tuning of the forcing parameters.

To characterize the temporal evolution of the patterns starting from
random initial conditions and to distinguish the resulting patterns
quantitatively we used the spectral entropy of the patterns, which
quantifies the effective number of Fourier modes of the patterns.
It allowed a clear distinction between patterns with three, four,
or more significant modes. However a more detailed quantitative characterization
of the transients that captures also the competition between multi-mode
structures like super-squares and anti-squares, which have the same
number of participating modes, is still an open problem (cf. \cite{RiMa06}).
The long-time scaling of the ordering process of such complex structures
and a comparison with the ordering in stripe \cite{HaAd00,BoVi01}
or hexagon patterns \cite{SaDe95} may also be interesting to study.
Most likely, the number of different types of defects and their mutual
interaction may play an important role. However, for the subharmonic
patterns discussed here the amplitude equations do not contain any
terms of even order. Therefore the strong interaction between defects
that is associated with those terms and that should make their dynamics
in particular interesting in the case of 5-mode patterns \cite{EcRi01}
will not be present here.

Here our focus was on systems below the Hopf bifurcation ($\mu<0)$.
It would be interesting to pursue a similar study above the Hopf bifurcation.
There the spontaneous oscillations, which do not arise in the Faraday
system, and their competition with phase-locked patterns driven by
the forcing may lead to additional complexity. For single-frequency
forcing above the Hopf bifurcation, labyrinthine stripe patterns arise
from the oscillations through front instabilities and stripe nucleation
\cite{YoHa02}. It is unknown what happens if the stripes are unstable
to the more complex patterns discussed here.

It would also be interesting to consider opposite signs for the detunings
$\nu_{21}$ and $\nu_{22}$ of the $1:2$-forcing,  frequency  so
that one of the detunings is just above the Hopf frequency, and the
other just below. Then only one of the two forcing terms will yield
a pattern-forming instability \cite{CoFr94}, while the other term
will induce a phase-locking of spatially homogeneous oscillations.
The result of the interaction between these two instabilities is not
known. 

Finally, we did not explore the  subcritical cases $b_{0}<0$ or $b(\theta)/b_{0}<-1$.
The weakly nonlinear analysis to cubic order is insufficient to make
predictions about  pattern selection in these cases.  Fig.\ref{fig:Cubic Coefs K is 1.857_phipio4}
suggests that for $K=2\cos(\tan^{-1}(2/5))$ large-amplitude
rectangle patterns with $\theta$ close to $\theta_{r}=\tan^{-1}(2/5)$
could be stabilized by the forcing

We gratefully acknowledge discussions with A. Rucklidge and M. Silber.
This work was supported by NSF grants DMS-322807 and DMS-0309657.

\appendix

\section*{Appendix\label{sec:Appendix}}

In Section \ref{sec:CGLE} we derived the form of the extended CGLE
(\ref{eq:cgle_final}) based on symmetry arguments. To illustrate
that the additional terms arise from a general forcing function in
a natural way we derive the coefficients of the equation here for
the Brusselator, which often has served as a simple model for chemical
oscillations \cite{LiHa00}. With the forcing included it can be written
as\begin{eqnarray}
\frac{\partial u}{\partial\tilde{t}} & = & 1-(1+B)u+D_{u}\Delta u+\left(1+f_{2}(\cos(\chi)\sin((2+\nu_{21})\tilde{t})+\sin(\chi)\sin((2+\nu_{22})\tilde{t}))+\right.\label{eq:brussel_u}\\
 &  & \left.+f_{1}\sin((1+\nu_{1})\tilde{t})+f_{3}\sin((3+\nu_{3})t)\right)u^{2}v,\nonumber \\
\frac{\partial v}{\partial\tilde{t}} & = & Bu+D_{v}\Delta v-u^{2}v,\label{eq:brussel_v}\end{eqnarray}
where $u$ and $v$ represent two reacting and diffusing chemical
species. In the formulation (\ref{eq:brussel_u},\ref{eq:brussel_v}),
which differs slightly from the form presented in \cite{LiHa00} in
the coefficients of the linear terms, the Hopf bifurcation occurs
at $B=2$ and the Hopf frequency is given by $\omega=1$. The small
parameters $\nu_{j}=O(\delta^{2})$, $\delta\ll1$, represent the
detuning (cf. eqs.(\ref{eq:detuning},\ref{eq:quasi-periodic})),
the parameters $f_{1}\equiv\delta^{3}\hat{f_{1}}$, $f_{2}\equiv\delta\hat{f_{2}}$,
and $f_{3}\equiv\delta\hat{f_{3}}$ represent the small 1:1-, 1:2-,
and 1:3 forcing strengths with $\hat{f}_{j}=\mathcal{O}(1)$, and
the parameters $D_{u}$ and $D_{v}$ are the diffusion coefficients.
To derive the extended CGLE (\ref{eq:cgle_final}) we expand (\ref{eq:brussel_u},\ref{eq:brussel_v})
near the Hopf bifurcation about the basic state $(u,v)=(1,B)$ as\[
\left(\begin{array}{c}
u\\
v\end{array}\right)=\left(\begin{array}{c}
1\\
B\end{array}\right)+\delta\left(\left(\begin{array}{c}
(-1-i)/2\\
1\end{array}\right)C(\tilde{x},\tilde{y},t)e^{i\omega\tilde{t}}+c.c.+\left(\begin{array}{c}
(-1-3i)/4\\
1/2\end{array}\right)\hat{f_{3}}e^{3i\omega\tilde{t}}+c.c.\right)+O(\delta^{2}).\]
To obtain the equation for the complex amplitude $A$ we extract the
frequency $\nu_{21}$ and write \begin{equation}
C=\sqrt{2/3}e^{i(\tan^{-1}(1/8)+\nu_{21}t)/2}A.\end{equation}
After rescaling the spatial coordinates as $(x,y)=\sqrt{(D_{u}+D_{v})/2}(\tilde{x},\tilde{y})$
 we arrive at Eq.(\ref{eq:cgle_final}) with the coefficients given
by \begin{eqnarray}
\mu & = & b/2+15|\hat{f_{3}}|^{2}/8,\\
\sigma & = & -(\nu_{21}/2+33|\hat{f_{3}}|^{2}/8),\\
\gamma & = & 6\hat{f_{2}}/\sqrt{65},\\
\rho & = & |\hat{f_{3}}|\sqrt{205/24},\\
\phi & = & \tan^{-1}(14/3)+\arg(\hat{f}_{3}).\end{eqnarray}
 As before $\nu=\nu_{22}-\nu_{21}$ and $\eta=\rho e^{i\phi}$.

To give an impression of the temporal evolution of the patterns as
they may be seen in experiments we show in (SimulationOver2Periods\_withFastOscs.mov)
(see also Fig.\ref{fig: u movie}) a movie of the $u$-component of
the Brusselator. More precisely, we show only the spatial dependence
associated with the wavevectors on the critical circle. In terms of
the expansion (\ref{eq:WNLansatz}) the corresponding amplitudes can
be written as 

\begin{align*}
u_{k}= & Z\,\left(\left(-\frac{1}{2}-\frac{i}{2}\right)\sqrt{\frac{2}{3}}\sum_{n=-\infty}^{\infty}(X_{n}+iY_{n})e^{in\nu t}e^{i\nu_{21}t/2+i\omega\tilde{t}}\right.\\
 & +\left.\left(-\frac{1}{2}+\frac{i}{2}\right)\sqrt{\frac{2}{3}}\sum_{n=-\infty}^{\infty}(X_{n}-iY_{n})e^{in\nu t}e^{-i\nu_{21}t/2-i\omega\tilde{t}}\right).\end{align*}

Here $Z$ represents the steady-state amplitude of the Fourier mode
given by the fixed-point solution of Eq.\ref{eq:GenAmpEqs}.

\bibliographystyle{unsrt}
\bibliography{/home/hermann/.bibfiles/journal}

\end{document}